\documentclass[%
 reprint,
 amsmath,amssymb,
 aps,
prl,superscriptaddress
]{revtex4-1}
\usepackage{mathtools}
\usepackage{braket}
\usepackage{siunitx}
\usepackage{graphicx}
\usepackage{dcolumn}
\usepackage{bm}
\usepackage[sort&compress]{natbib}
\usepackage{subfigure}
\usepackage{float}
\usepackage{ulem}

\begin{document}
\bibliographystyle{apsrev4-1}
\preprint{APS/123-QED}

\title{Nonlinear resonant x-ray Raman scattering}
\author{Johann Haber}
\email{jhaber@stanford.edu}
\affiliation{Stanford PULSE Institute, SLAC National Accelerator Laboratory, Menlo Park, Californa 94025, USA}
\author{Andreas Kaldun}
\affiliation{Stanford PULSE Institute, SLAC National Accelerator Laboratory, Menlo Park, Californa 94025, USA}
\author{Samuel W. Teitelbaum}
\altaffiliation[Current address: ]{Department of Physics, Arizona State University, Tempe, Arizona  85287, USA}
\affiliation{Stanford PULSE Institute, SLAC National Accelerator Laboratory, Menlo Park, Californa 94025, USA}
\author{Alfred Q.R. Baron}
\affiliation{Riken SPring-8 Center, Kouto 1-1-1 Sayo, Hyogo 679-5148, Japan}
\author{Philip H. Bucksbaum}
\affiliation{Stanford PULSE Institute, SLAC National Accelerator Laboratory, Menlo Park, Californa 94025, USA}
\author{Matthias Fuchs}
\affiliation{Department of Physics and Astronomy, University of Nebraska, Lincoln, Nebraska 68588, USA}
\author{Jerome B. Hastings}
\affiliation{Stanford PULSE Institute, SLAC National Accelerator Laboratory, Menlo Park, Californa 94025, USA}
\author{Ichiro Inoue}
\affiliation{Riken SPring-8 Center, Kouto 1-1-1 Sayo, Hyogo 679-5148, Japan}
\author{Yuichi Inubushi}
\affiliation{Riken SPring-8 Center, Kouto 1-1-1 Sayo, Hyogo 679-5148, Japan}
\affiliation{Japan Synchrotron Radiation Research Institute, Kouto 1-1-1 Sayo, Hyogo 679-5198, Japan}
\author{Dietrich Krebs}
\affiliation{Department of Physics, Universit\"at Hamburg, Jungiusstrasse 9, 20355 Hamburg, Germany}
\affiliation{Center for Free-Electron Laser Science, Deutsches Elektronen-Synchrotron DESY, Notkestrasse 85, 22607, Hamburg, Germany}
\affiliation{Max Planck School of Photonics, Friedrich-Schiller-University of Jena, Albert-Einstein-Strasse 6, 07745 Jena, Germany}
\author{Taito Osaka}
\affiliation{Riken SPring-8 Center, Kouto 1-1-1 Sayo, Hyogo 679-5148, Japan}
\author{Robin Santra}
\affiliation{Center for Free-Electron Laser Science, Deutsches Elektronen-Synchrotron DESY, Notkestrasse 85, 22607, Hamburg, Germany}
\affiliation{Department of Physics, Universit\"at Hamburg, Jungiusstrasse 9, 20355 Hamburg, Germany}
\author{Sharon Shwartz}
\affiliation{Physics Department and Institute of Nanotechnology, Bar Ilan University, Ramat Gan 52900, Israel}
\author{Kenji Tamasaku}
\affiliation{Riken SPring-8 Center, Kouto 1-1-1 Sayo, Hyogo 679-5148, Japan}
\author{David A. Reis}
\email{dreis@stanford.edu}
\affiliation{Stanford PULSE Institute, SLAC National Accelerator Laboratory, Menlo Park, Californa 94025, USA}

\date{\today}

\begin{abstract}
We report the observation of a novel nonlinear effect in the hard x-ray range. Upon illuminating Fe and Cu metal foils with intense x-ray pulses tuned near their respective K edges, photons at nearly twice the incoming photon energy are emitted. The signal  rises quadratically with the incoming intensity, consistent with two-photon excitation. 
The spectrum of emitted high-energy photons comprises multiple Raman lines that disperse with the incident photon energy. Upon reaching the double K-shell ionization threshold, the signal strength undergoes a marked rise. Above this threshold, the lines cease dispersing, turning into florescence lines with energies much greater than obtainable by single electron transitions, and additional Raman lines appear. We attribute these processes  to  electron-correlation mediated multielectron transitions involving double-core hole excitation and various two-electron de-excitation processes to a final state involving one or more L and M core-holes. 
\end{abstract}

\pacs{Valid PACS appear here}
\maketitle

The extremely high peak-brightness of x-ray free-
electron lasers enables nonlinear resonant interactions
to be driven on time-scales shorter than typical femtosecond-scale core-hole 
lifetimes. A number of nonlinear phenomena have been reported in the gas phase, including Rabi-oscillations~\cite{kanter} and multiple sequential ionization~\cite{Young2010,Gerken2017, Rudenko2017} as well as both single-site~\cite{Tamasaku2013} and double-site double~\cite{Fang2010} core-hole formation.  Other effects, such as two-photon absorption,~\cite{ghimi,Tamasaku2014,Ghimire2016} stimulated Raman scattering,~\cite{Weninger2013,Yoneda,kroll}  and lasing~\cite{rohringer,Yoneda} have been reported in both gas and condensed phase.  Non-resonant, non-sequential effects have also been reported in solids including  phase-matched second harmonic generation~\cite{shwartz1}, and two-photon Compton scattering~\cite{compton}.  In the latter case, the simultaneous inelastic scattering  of two high-energy photons into a high-energy photon showed an anomalously large red-shift from the second harmonic compared to expectations that it should behave like scattering from free-electrons far from resonance  \cite{Krebs2019,Venkatesh2020}. In addition to their fundamental importance, these phenomena open the possibility of novel spectroscopies that combine structural information and atomic specificity in ways that are inaccessible for single photon interactions.  

Here we report resonant two-photon inelastic x-ray scattering from two different cubic polycrystalline  metallic foils:  6-fold coordinated (bcc) 3d ferromagnetic Fe and 12-fold coordinated (fcc) monovalent Cu.  In each case we
excite with photons of energy from just below the relevant single K-shell ionization threshold to well above the
sequential double K-shell threshold, and measure higher energy photons below twice the input energy. Both metals show qualitatively similar results that we interpret in terms of their atomic-like nonlinear response, as we do not expect to resolve band structure effects in the current experiment.  We find the scattered high-energy photon signal scales quadratically with the input intensity, consistent with a perturbative two-photon absorption process.  The spectrum of emitted high-energy photons show both dispersive and non-dispersive lines that we attribute to double-core hole mediated resonant x-ray Raman and two-electron one photon fluorescence (TEOP~\cite{Hoszowska2011}). The former is somewhat analogous to hyper-Raman scattering~\cite{Terhune1965} although in our case the inelastic shift disperses with the incident photon energy rather than twice the energy.  This can be considered as an anti-Stokes process from a core-excited intermediate state where the excess energy is taken up by one or more photo-electrons and lines. In all cases both K shell electrons are excited and subsequently de-excited, in the process.

\begin{figure}
\includegraphics[scale=.19]{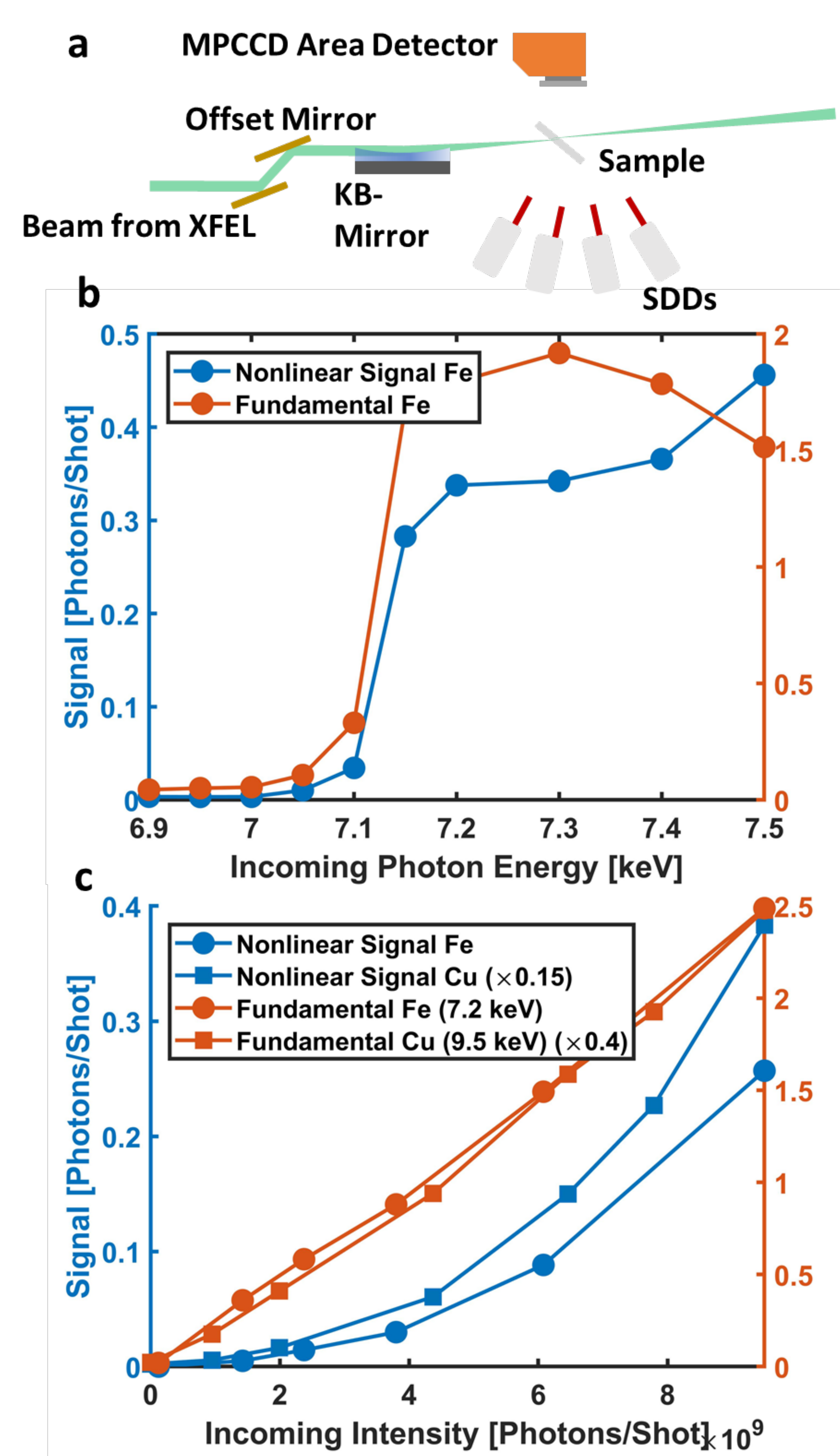}
\caption{(a) Setup of the experiment. An intense x-ray beam focused by a Kirkpatrick-Baez (KB) mirror pair illuminates a thin Fe or Cu foil. An area pixelated detector (MPCCD) and four silicon drift detectors (SDDs) detect higher energy photons scattered by the sample.
(b) Average number of photons detected on the MPCCD per shot for the iron foil as a function of incident x-ray energy near the K-egde (7.11 keV) around the fundamental, close to the K$\alpha$ line (red), and at higher energies, close to the second harmonic (blue).
%
(c) Intensity dependence of scattered lower-energy (red) and higher-energy signal (blue) on the MPCCD for Fe for incident energies of 7.2 keV for Fe (circles) and 9.5 keV Cu (squares). Errorbars are smaller than symbols.}
\label{fig:setup}
\end{figure}
The experiment was performed at Beamline 3 of the SACLA x-ray free electron laser at the RIKEN institute in Japan~\cite{Tono2013}. The experimental setup is sketched in Fig.~\ref{fig:setup}. The pulse, of bandwidth $\approx 30$ eV and an estimated $10$ fs length is focused to $1.5~\mathrm{\mu m} \times 1.5~\mathrm{\mu m}$ (FWHM). The samples, Fe and Cu foils of thickness $5 ~\mathrm{\mu m}$ (on the order of 1-2 absorption lengths), are placed at the focus position.  Two sets of harmonic rejection mirrors were used with either 7.5 keV or 15 keV cutoff depending on the incident energy. 
The average number of photons per pulse was up to $\sim 10^{10}$ in the fundamental with contamination of up to a few hundred photons at the undulator second harmonic at the sample location at the  Fe K-edge. The resulting intensity per pulse is $\sim 10^{17}\text{W}/\text{cm}^2$, and the fluence $\sim 5\times 10^{17}\text{~photons}/\text{cm}^2$. The fluence is sufficiently low that we do not saturate the K-shell absorption, yet high enough that the sample is destroyed in a single pulse; the foils hence are continuously scanned, such that each shot  illuminates a fresh spot on the sample. We use two sets of detectors, the first one being a Multiport CCD (MPCCD) with $256\times 512$ pixels of $50~\mathrm{\mu m} \times 50~\mathrm{\mu m}$ size. It was positioned next to the sample at an angle of $90^{\circ}$ to the beam direction in the polarization plane, with a distance of $20 $ cm from the focus. The MPCCD has a very rough energy resolution of about $1$ keV. 
For enhanced energy resolution and a more detailed insight into the physical origins of the signal, we use four AMPTEK silicon drift detectors (SDDs) opposite the MPCCD placed in $30^{\circ}$ intervals (in the polarization plane). The distance of the detectors from the sample position is $20$ cm, yielding a smaller solid angle than the entirety of the MPCCD. The energy resolution is $200$ eV around $14$ keV. Both the MPCCD and the SDDs were heavily shielded with aluminum to limit pile-up. From the fundamental signal of $2.5$ photons/shot spread over $256 \times 512$ pixels,  we estimate the number of pile-up events to be $\sim 10^{-5}$ per shot for the MPCCD, well below the nonlinear signal. The pile-up  of the SDD is at least an order of magnitude below the relevant signal.\\

Figure ~\ref{fig:setup} (b) shows the average number of photons detected by the MPCCD  per shot for the Fe foil as a function of incoming photon energy from about 200 eV below to 400 eV above the K-edge at 7.11 keV.  The red  curve represents the average signal with energy near the fundamental, close to the Fe K$\alpha$- and K$\beta$-lines and the blue curve shows the average signal with energy near the second harmonic of the incident beam. For simplicity we will refer to these as the low and high energy signals. 
 The Al shielding suppresses the Fe K$\alpha$ by $7-8$ orders of magnitude, while transmitting about $3-6\%$ in the higher energy area of interest. 
Both the low and the higher energy signals rise sharply as we scan across the K-edge. While the low energy signal slowly decreases with increasing photon energy, the higher energy signal shows a secondary rise as the input is tuned above $7.4$ keV.
 This corresponds to the double K shell ionization (DKI) threshhold~\cite{Hoszowska2009}, i.e. the energy at which the second electron of the Fe $1s$ shell can be ionized (at $7.45$ keV by Hartree Fock calculations~\cite{cowan}). Its threshold is higher than the K-edge due to
the stronger binding of the remaining 1s-electron to the unscreened ionic core.

 We attenuate the incident beam with Al filters to measure the intensity dependence of the signal. The results are shown in Fig.~\ref{fig:setup} (c) for both Fe at 7.2 keV (circles) and Cu at 9.5 keV (squares), well above its K-edge (8.98 keV). In each case, we display the low energy signal in red and the high-energy signal in blue.   
  In both Fe and Cu the low-energy signal varies linearly with the incoming intensity, and  higher energy signal varies quadratically with incoming intensity of the fundamental.  Since the filters have negligible absorption at the undulator second harmonic, and we can similarly rule out pile-up, this indicates that the higher-energy signal involves two-photon excitation. 

\begin{figure}[t]
\includegraphics[scale=.23]{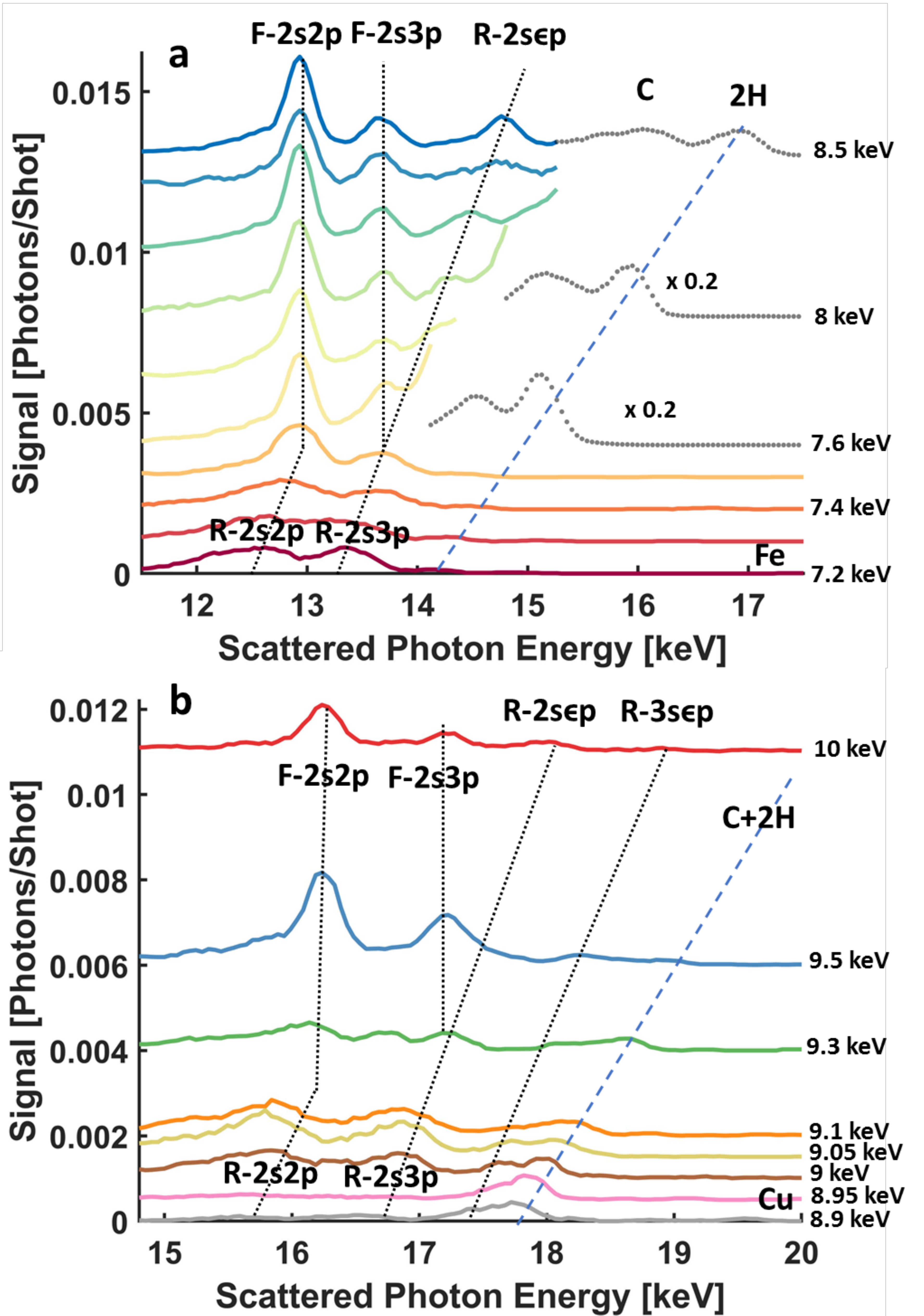}
\caption{Emission spectra for (a) Fe, and (b) Cu measured with the SDDs for different incident photon energies (indicated on the right), vertically displaced for clarity.  
The dashed lines are a guide to the eye. The labels starting with F or R correspond to the various nonlinear processes described in the text. The higher energy features labeled 2H and C are dominated by contamination from elastic scattering and Compton scattering of the residual undulator second harmonic.  }
\label{fig:disp}
\end{figure}

To resolve better the spectra of the emitted photons, we turn to the data taken with the SDDs. The spectra around the higher emitted energies are shown in Fig.~\ref{fig:disp}(a) for Fe and (b) for Cu. Both metals display a rich set of features. For Fe excited above the K edge two peaks appear at $12.6$ and $13.3$ keV. As the incoming photon energy is increased, these peaks shift to higher energies. This proceeds until the incident x-ray photon energy reaches the DKI threshold beyond $7.45$ keV. At this point the two lines become nondispersive which is indicative of fluorescence. They are however centered at $12.9$ and $13.65$ keV respectively, considerably higher than the binding energy for a single electron.
In addition, a smaller, less intense line seems to branch off from the $13.65$ keV line, again dispersing with the incident photon energy. \\

\begin{figure}[t!]
\includegraphics[scale=.25]{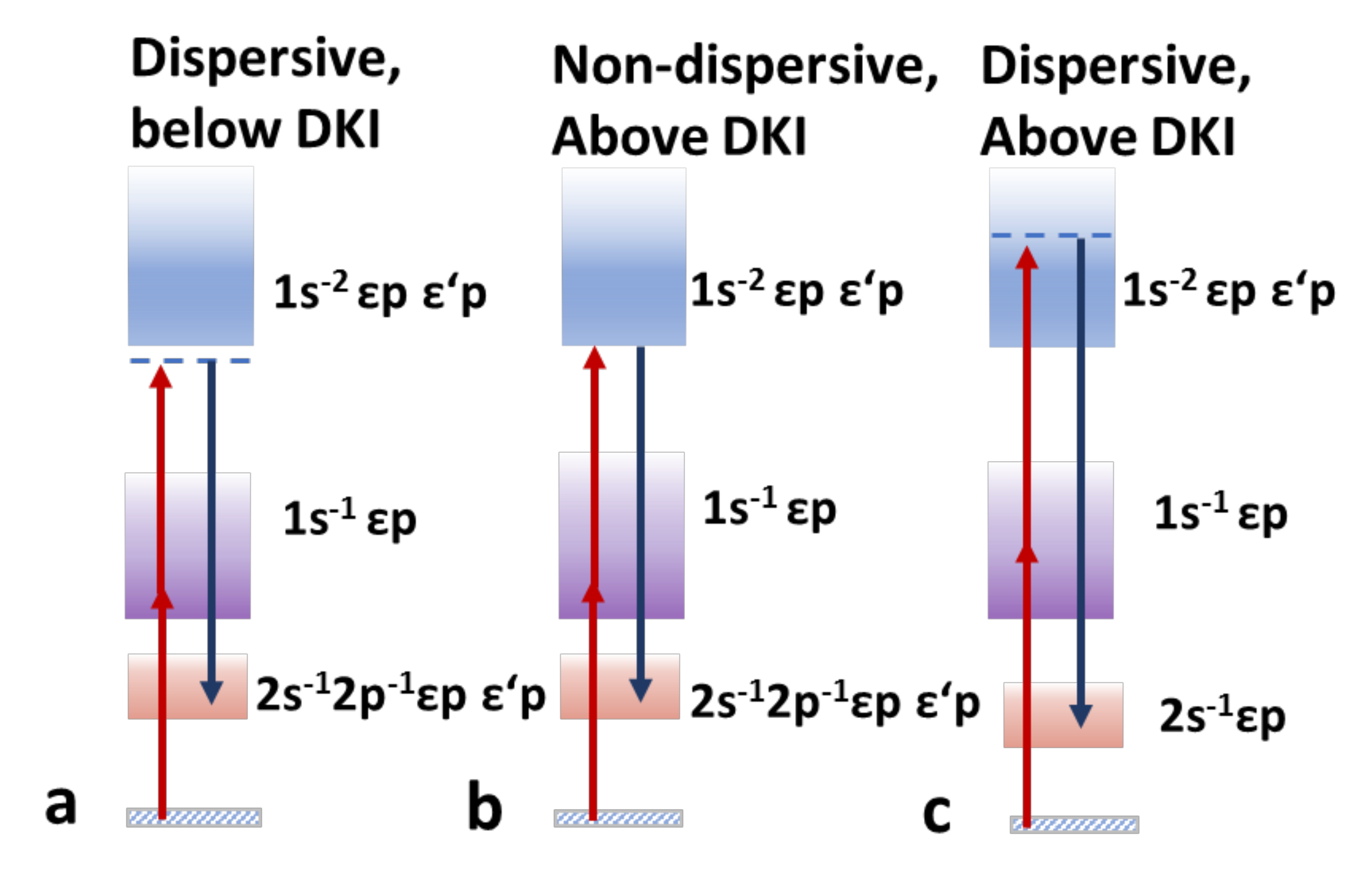}
\caption{In the R-$2s2p$ process (a) an x-ray photon ejects an electron from the 1s shell into the continuum with energy $\epsilon$. A second $1s$ electron is then excited virtually. Inner shell electrons decay into the $1s$ holes and emit a single photon that disperses with the incident energy. Energy conservation dictates that the emitted photon energy is less than the energy difference between the$2s^{-1}2p^{-1}$ and $1s^{-2}$ double core-hole configurations (with the photoelectrons taken up the balance). In F-$2s2p$ (b) above the DKI threshold  the second photon has enough energy to eject the second 1s electron, and a fluorescence photon is emitted. The energy corresponds to the energy difference between the double core configurations. In the R-$2s\epsilon$p process (c) the TEOP transition proceeds by the concerted decay of one continuum eletron and on core electron. Again, the energy of the emitted photon disperses with the photon energy. In Fig.~\ref{fig:disp}, this type of transition is labeled . }
\label{fig:conf}
\end{figure}

The higher energy spectra for copper shown in Fig.~\ref{fig:disp} (b) are qualitatively similar to iron. Below the double K-shell ionization threshold, which for copper is some $400$ eV above the K edge, at $9.4$ keV, we observe three dispersive lines. Above the threshold, we again observe two fluorescence lines with energy well above the $1s$ binding energy, as well as two smaller dispersive lines at higher energies. In both Fe and Cu elastic (2H) and Compton (C) scattering of the residual undulator second harmonic appears at the high energy end of the spectra, especially when excited above 7.5 keV, above which we used the SACLA $15$ keV mirrors which have limited rejection of the harmonic near the cut-off.
We interpret the nondispersive line above the DKI threshold as concerted two-electron, one-photon (TEOP) transitions that simultaneously fill a doubly ionized K-shell mediated by electron-electron correlations ~\cite{Kelly1976, Aberg1976, Hoogkamer1976, Hodge1977}. 
Similar lines were  observed in the 1970s in experiments on heavy ion bombardment of Fe foils~\cite{Wolfli1975, Stoller1977}.  
Relativistic Dirac-Hartree-Fock calculations give the average energy difference between the $\ket{1s^{-2}}$ and the $\ket{2s^{-1}2p^{-1}}$ or $\ket{2s^{-1}3p^{-1}}$ configurations for Fe to be $12.98$ and $13.67$ keV, in agreement (within the detector resolution) with our (Fig.~\ref{fig:disp}(a)) as well as previous data~\cite{Hodge1977}. Here  $\ket{ns^{-x}mp^{-x}}$ denotes a configuration in which $x$ electrons are missing in shells with the principal quantum numbers $n$ and $m$. $s$ and $p$ correspond to angular momentum quantum numbers $l =0,1$ respectively. \\
In our case the double core hole excitation stems from two photon excitation of two $1s$ electrons into the continuum, leaving them with energies $\epsilon$ and $\epsilon' $ and (predominantly) angular momentum $l=1$. In both cases radiative relaxation will occasionally occur by the $2s$ and $2p$  or $2s$ and $3p$ electrons respectively decaying into the empty $1$s shell by a process involving direct Coulomb interaction with each other or other electrons. In doing so they emit a photon with constant energy $E_x = E_{np} + E_{ms} - E_K - E_{KK}$. $E_{np} (E_{ms})$ are the binding energies of the electrons with the principal quantum numbers $n$ ($m$) and the angular momentum quantum numbers $l=1~(0)$. $E_K$ and $E_{KK}$ are the K edge energy and DKI threshold energy. The process corresponds to the lines labeled with F-$nsmp$ in Fig.~\ref{fig:disp} and is depicted   in Fig.~\ref{fig:conf} (b).

The continuum electrons remain spectators in the TEOP process described above, i.e. they do not contribute further, except for some Coulomb interactions~\cite{Kelly1976}.\\ 
We attribute the dispersive lines that appear below the DKI threshold (R-$nsmp$ in Fig.~\ref{fig:disp}) to a two-photon Raman process analogous to the single-photon process seen for excitation below the K-edge~\cite{Eisenberger1976}.  In our case, the second x-ray photon is of insufficient energy to make a real transition to the continuum --- but it may excite a virtual continuum electron, thus creating a virtual hollow atom, as indicated in Fig.~\ref{fig:conf} (a). The $2s$ and $2p$ or $2s$ and $3p$ electrons may still decay into the $1s$ shell, as in TEOP fluorescence, but the energy of the photon they emit has to be smaller than the total energy difference between the configurations by energy conservation. The dispersion is  given as $E_{x} = E_{np} + E_{2s} + E_{\omega} - E_{K}$ where $E_{\omega}$ is the energy of the incident x-ray photon. This sub-threshold TEOP de-excitation process can be viewed as a nonlinear, two-photon (in), two-electron Raman transition. The scattered photon disperses with the energy of a single incident photon and the excess energy is taken up by one of the photoelectrons. In this process the continuum electrons also remain spectators. In principle, however, they can become participators. Assuming that  excitations into the continut eneruum are dipolar, the continuum electrons are $p$ electrons. Symmetry  requires the two-electron transitions to encompass both $s$ and $p$ electrons~\cite{Goudsmit1931}. A TEOP transition involving the $p$-continuum electron of energy $\epsilon'$ and the $2s$ electron should result in an emitted photon with the energy $E_x = E_{2s} - E_{KK} - E_{K} + \epsilon'$. Of course, $\epsilon' - E_{KK} = E_{\omega}$, leading to dispersion with the incident photon energy, such as we saw below the DKI threshold. The energy of the photon $E_x$ that results from it will be larger than that of the nondispersive lines. This process could explain the origin of the dispersive lines  above the DKI (R-$ns\epsilon$p).

\begin{figure}
\includegraphics[scale=.23]{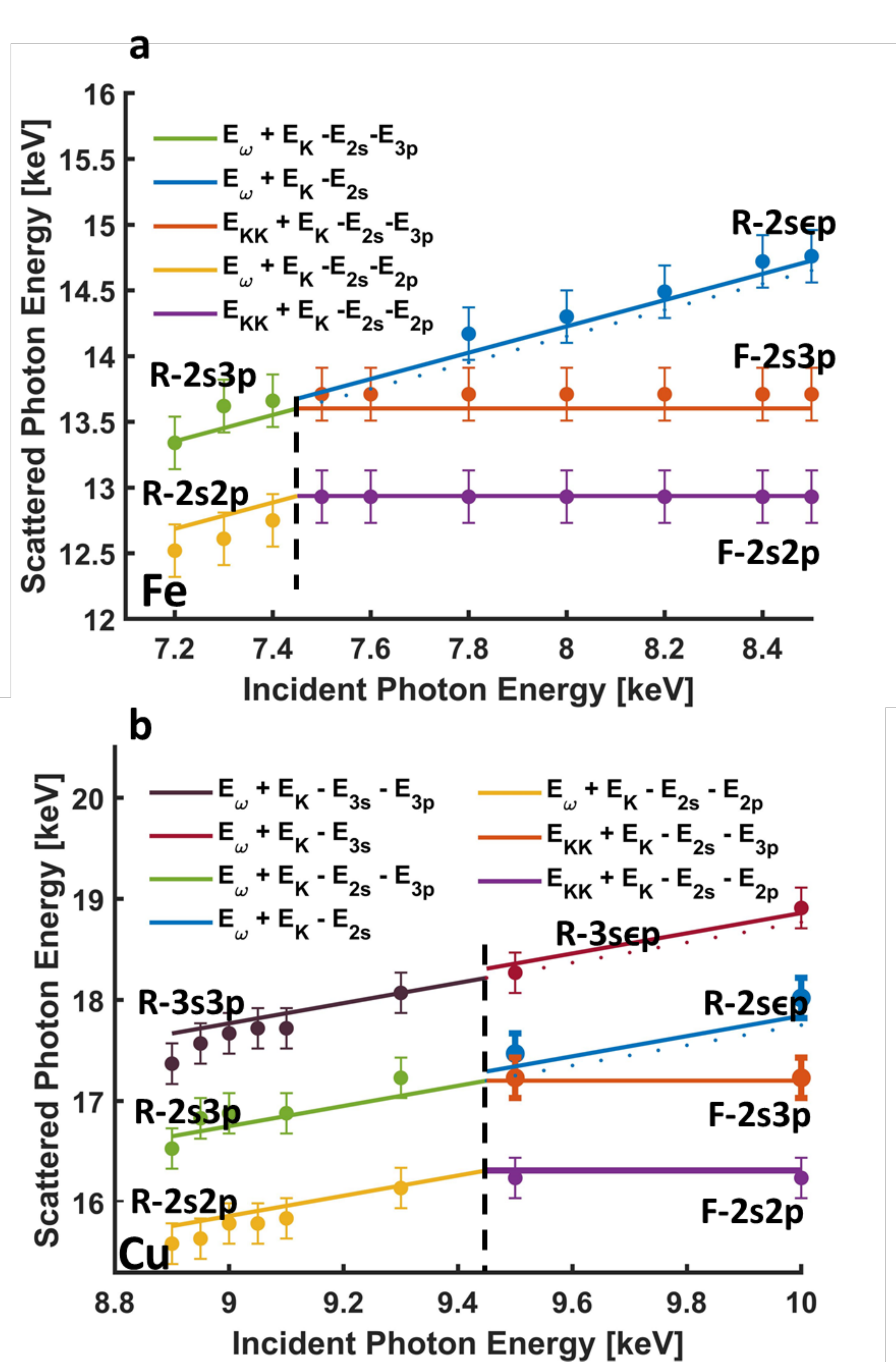}
\caption{Dispersion of the measured signal peaks for a) Fe and b) Cu. Dots indicate data points extracted from the spectra in Fig.~\ref{fig:disp}. The approximate systematic error corresponds to  the detector resolution of $200$ eV,. The solid lines are dispersion curves of TEOP transitions, the shells involved indicated in the legend. The dashed grey line indicates the DKI threshold.}
\label{fig:rd}
\end{figure}

We plot the dispersion relations for the three processes given above against the centers of mass of the spectral peaks in Fig.~\ref{fig:rd} (a) and (b) for Fe and Cu respectively. The binding energies were calculated by the Hartree Fock method with relativistic corrections.  Within the SDD resolution of $\sim$200 eV, the expected dispersion  agrees nicely with the data.
 
We estimate  the effective two-photon absorption cross-section  double core-hole generation in Fe using the measured F-$2s2p$ lines and compare it to a rough calculation for the sequential process.
The experimental estimate is similar to the approach used for two-photon one-electron absorption in Zr~\cite{Ghimire2016} assuming a Gaussian x-ray pulse in space and time with the focal area of $2.5 ~\mu\mathrm{m}^2$ ($\pi/2 (FWHM/\sqrt{2log2})^2)$) and a pulse length of $10$ fs. 
The estimate yields an effective generalized absorption cross-section $\sigma^{(2),\text{exp}}_\text{DCH} \approx 2.7 \times 10^{-53}~\mathrm{cm^4 s}$, assuming isotropic emission, for a total signal of $\sim 0.015$ photons/shot at $7.5$ keV (obtained by integrating over the $\ket{1s^{-2}} \rightarrow \ket{2s^{-1}2p^{-1}}$ (F-$2s2p$) transition line) for  $9.5\times 10^{9}$ incident photons, a detector quantum efficiency, $Q_E = 0.04$ (including Al shielding). Furthermore, we take the calculated branching ratio of double to single electron fluoresence $2\times10^{-4}$ from~\cite{Kelly1976} as a higher bound for TEOP fluorescent yield. The uncertainty in $\sigma^{(2),\text{exp}}_\text{DCH}$ is dominated by statistical uncertainties in the intensity distribution of the focused x-ray pulse as well as uncertainties in the theoretical calculations for the branching ratio, which easily amount to an order of magnitude.

%
%
%
We obtain a theoretical estimate of $\sigma^{(2),\text{th}}_\text{DCH} = \frac{1}{2} \sigma_K \tau_{1s} \sigma_K \approx 2.9 \times 10^{-55}\mathrm{cm^4s}$, where $\sigma_K = 3.33 \times 10^{-20}~\mathrm{cm^2}$ is the absorption cross-section for the neutral Fe at 7.5 keV~\cite{xcom}, the  single core-hole lifetime $\tau_{1s} \sim 0.52$ fs  and we include a factor of $1/2$ to account (approximately) for the reduced cross-section of the $1s^{-1}$ state due to the presence of a single electron in the K-shell. This is a factor of 90 smaller than the experimental cross section. Additionally, the rough theoretical estimate does not consider the details of the intermediate states or the random temporal structure of the SASE pulse~\cite{Tamasaku2013}.
For comparison the experimental (theoretical) estimates for the sequential double-core cross-section is about  6 (4) orders of magnitude higher than that for the theoretical concerted two-photon single K absorption of Fe at threshold \cite{Ghimire2016}.


We note that more accurate measurements of branching ratios and fluorescence photon energies of multi-core hole atoms could be used to benchmark many-body and relativistic descriptions for atomic physics~\cite{Hoszowska2011,Deutsch1992}. Also, while the current experiments were performed at high excitation densities that led to permanent damage of the sample, the combination of higher resolution afforded by transition edge sensors~\cite{Doriese2017} and the higher repetition rate of future FELs~\cite{LCLS2} point to the possibility for spectroscopic studies in the condensed phase. The sensitivity to valence electrons afforded by such studies could prove complementary to RIXS~\cite{RIXSRMP}, especially in two-color excitations where one photon is resonant near a transition-metal L-edge and the second is well-above the edge.  This would allow for combined elemental specificity of low Z L-edges with atomic-scale momentum resolution that is not limited by the low energy of the L-edge. 
\begin{acknowledgments}
This research was primarily supported by the AMOS program within the Chemical Sciences Division of the Office of Basic Energy Sciences, Office of Science, U.S. Department of Energy. The XFEL experiments were performed at the BL3 of SACLA with the approval of the Japan Synchrotron Radiation Research Institute (JASRI) (Proposal Nos. 2017A8038, 2018A8069 and 2019A8078).  DAR acknowledges discussions with Steve Southworth and Linda Young.
\end{acknowledgments}

\bibliography{mybib}

\begin{thebibliography}{35}%
\makeatletter
\providecommand \@ifxundefined [1]{%
 \@ifx{#1\undefined}
}%
\providecommand \@ifnum [1]{%
 \ifnum #1\expandafter \@firstoftwo
 \else \expandafter \@secondoftwo
 \fi
}%
\providecommand \@ifx [1]{%
 \ifx #1\expandafter \@firstoftwo
 \else \expandafter \@secondoftwo
 \fi
}%
\providecommand \natexlab [1]{#1}%
\providecommand \enquote  [1]{``#1''}%
\providecommand \bibnamefont  [1]{#1}%
\providecommand \bibfnamefont [1]{#1}%
\providecommand \citenamefont [1]{#1}%
\providecommand \href@noop [0]{\@secondoftwo}%
\providecommand \href [0]{\begingroup \@sanitize@url \@href}%
\providecommand \@href[1]{\@@startlink{#1}\@@href}%
\providecommand \@@href[1]{\endgroup#1\@@endlink}%
\providecommand \@sanitize@url [0]{\catcode `\\12\catcode `\$12\catcode
  `\&12\catcode `\#12\catcode `\^12\catcode `\_12\catcode `\%12\relax}%
\providecommand \@@startlink[1]{}%
\providecommand \@@endlink[0]{}%
\providecommand \url  [0]{\begingroup\@sanitize@url \@url }%
\providecommand \@url [1]{\endgroup\@href {#1}{\urlprefix }}%
\providecommand \urlprefix  [0]{URL }%
\providecommand \Eprint [0]{\href }%
\providecommand \doibase [0]{http://dx.doi.org/}%
\providecommand \selectlanguage [0]{\@gobble}%
\providecommand \bibinfo  [0]{\@secondoftwo}%
\providecommand \bibfield  [0]{\@secondoftwo}%
\providecommand \translation [1]{[#1]}%
\providecommand \BibitemOpen [0]{}%
\providecommand \bibitemStop [0]{}%
\providecommand \bibitemNoStop [0]{.\EOS\space}%
\providecommand \EOS [0]{\spacefactor3000\relax}%
\providecommand \BibitemShut  [1]{\csname bibitem#1\endcsname}%
\let\auto@bib@innerbib\@empty
\bibitem [{\citenamefont {Kanter}\ \emph {et~al.}(2011)\citenamefont {Kanter},
  \citenamefont {Kr{\"{a}}ssig}, \citenamefont {Li}, \citenamefont {March},
  \citenamefont {Ho}, \citenamefont {Rohringer}, \citenamefont {Santra},
  \citenamefont {Southworth}, \citenamefont {DiMauro}, \citenamefont {Doumy},
  \citenamefont {Roedig}, \citenamefont {Berrah}, \citenamefont {Fang},
  \citenamefont {Hoener}, \citenamefont {Bucksbaum}, \citenamefont {Ghimire},
  \citenamefont {Reis}, \citenamefont {Bozek}, \citenamefont {Bostedt},
  \citenamefont {Messerschmidt},\ and\ \citenamefont {Young}}]{kanter}%
  \BibitemOpen
  \bibfield  {author} {\bibinfo {author} {\bibfnamefont {E.~P.}\ \bibnamefont
  {Kanter}}, \bibinfo {author} {\bibfnamefont {B.}~\bibnamefont
  {Kr{\"{a}}ssig}}, \bibinfo {author} {\bibfnamefont {Y.}~\bibnamefont {Li}},
  \bibinfo {author} {\bibfnamefont {A.~M.}\ \bibnamefont {March}}, \bibinfo
  {author} {\bibfnamefont {P.}~\bibnamefont {Ho}}, \bibinfo {author}
  {\bibfnamefont {N.}~\bibnamefont {Rohringer}}, \bibinfo {author}
  {\bibfnamefont {R.}~\bibnamefont {Santra}}, \bibinfo {author} {\bibfnamefont
  {S.~H.}\ \bibnamefont {Southworth}}, \bibinfo {author} {\bibfnamefont
  {L.~F.}\ \bibnamefont {DiMauro}}, \bibinfo {author} {\bibfnamefont
  {G.}~\bibnamefont {Doumy}}, \bibinfo {author} {\bibfnamefont {C.~A.}\
  \bibnamefont {Roedig}}, \bibinfo {author} {\bibfnamefont {N.}~\bibnamefont
  {Berrah}}, \bibinfo {author} {\bibfnamefont {L.}~\bibnamefont {Fang}},
  \bibinfo {author} {\bibfnamefont {M.}~\bibnamefont {Hoener}}, \bibinfo
  {author} {\bibfnamefont {P.~H.}\ \bibnamefont {Bucksbaum}}, \bibinfo {author}
  {\bibfnamefont {S.}~\bibnamefont {Ghimire}}, \bibinfo {author} {\bibfnamefont
  {D.~A.}\ \bibnamefont {Reis}}, \bibinfo {author} {\bibfnamefont {J.~D.}\
  \bibnamefont {Bozek}}, \bibinfo {author} {\bibfnamefont {C.}~\bibnamefont
  {Bostedt}}, \bibinfo {author} {\bibfnamefont {M.}~\bibnamefont
  {Messerschmidt}}, \ and\ \bibinfo {author} {\bibfnamefont {L.}~\bibnamefont
  {Young}},\ }\href {\doibase 10.1103/PhysRevLett.107.233001} {\bibfield
  {journal} {\bibinfo  {journal} {Phys. Rev. Lett.}\ }\textbf {\bibinfo
  {volume} {107}},\ \bibinfo {pages} {233001} (\bibinfo {year}
  {2011})}\BibitemShut {NoStop}%
\bibitem [{\citenamefont {Young}\ \emph {et~al.}(2010)\citenamefont {Young},
  \citenamefont {Kanter}, \citenamefont {Kr{\"{a}}ssig}, \citenamefont {Li},
  \citenamefont {March}, \citenamefont {Pratt}, \citenamefont {Santra},
  \citenamefont {Southworth}, \citenamefont {Rohringer}, \citenamefont
  {DiMauro}, \citenamefont {Doumy}, \citenamefont {Roedig}, \citenamefont
  {Berrah}, \citenamefont {Fang}, \citenamefont {Hoener}, \citenamefont
  {Bucksbaum}, \citenamefont {Cryan}, \citenamefont {Ghimire}, \citenamefont
  {Glownia}, \citenamefont {Reis}, \citenamefont {Bozek}, \citenamefont
  {Bostedt},\ and\ \citenamefont {Messerschmidt}}]{Young2010}%
  \BibitemOpen
  \bibfield  {author} {\bibinfo {author} {\bibfnamefont {L.}~\bibnamefont
  {Young}}, \bibinfo {author} {\bibfnamefont {E.~P.}\ \bibnamefont {Kanter}},
  \bibinfo {author} {\bibfnamefont {B.}~\bibnamefont {Kr{\"{a}}ssig}}, \bibinfo
  {author} {\bibfnamefont {Y.}~\bibnamefont {Li}}, \bibinfo {author}
  {\bibfnamefont {A.~M.}\ \bibnamefont {March}}, \bibinfo {author}
  {\bibfnamefont {S.~T.}\ \bibnamefont {Pratt}}, \bibinfo {author}
  {\bibfnamefont {R.}~\bibnamefont {Santra}}, \bibinfo {author} {\bibfnamefont
  {S.~H.}\ \bibnamefont {Southworth}}, \bibinfo {author} {\bibfnamefont
  {N.}~\bibnamefont {Rohringer}}, \bibinfo {author} {\bibfnamefont {L.~F.}\
  \bibnamefont {DiMauro}}, \bibinfo {author} {\bibfnamefont {G.}~\bibnamefont
  {Doumy}}, \bibinfo {author} {\bibfnamefont {C.~A.}\ \bibnamefont {Roedig}},
  \bibinfo {author} {\bibfnamefont {N.}~\bibnamefont {Berrah}}, \bibinfo
  {author} {\bibfnamefont {L.}~\bibnamefont {Fang}}, \bibinfo {author}
  {\bibfnamefont {M.}~\bibnamefont {Hoener}}, \bibinfo {author} {\bibfnamefont
  {P.~H.}\ \bibnamefont {Bucksbaum}}, \bibinfo {author} {\bibfnamefont {J.~P.}\
  \bibnamefont {Cryan}}, \bibinfo {author} {\bibfnamefont {S.}~\bibnamefont
  {Ghimire}}, \bibinfo {author} {\bibfnamefont {J.~M.}\ \bibnamefont
  {Glownia}}, \bibinfo {author} {\bibfnamefont {D.~A.}\ \bibnamefont {Reis}},
  \bibinfo {author} {\bibfnamefont {J.~D.}\ \bibnamefont {Bozek}}, \bibinfo
  {author} {\bibfnamefont {C.}~\bibnamefont {Bostedt}}, \ and\ \bibinfo
  {author} {\bibfnamefont {M.}~\bibnamefont {Messerschmidt}},\ }\href {\doibase
  10.1038/nature09177} {\bibfield  {journal} {\bibinfo  {journal} {Nature}\
  }\textbf {\bibinfo {volume} {466}},\ \bibinfo {pages} {56} (\bibinfo {year}
  {2010})}\BibitemShut {NoStop}%
\bibitem [{\citenamefont {Gerken}\ \emph {et~al.}(2014)\citenamefont {Gerken},
  \citenamefont {Klumpp}, \citenamefont {Sorokin}, \citenamefont {Tiedtke},
  \citenamefont {Richter}, \citenamefont {B{\"{u}}rk}, \citenamefont {Mertens},
  \citenamefont {Jurani\'{c}},\ and\ \citenamefont {Martins}}]{Gerken2017}%
  \BibitemOpen
  \bibfield  {author} {\bibinfo {author} {\bibfnamefont {N.}~\bibnamefont
  {Gerken}}, \bibinfo {author} {\bibfnamefont {S.}~\bibnamefont {Klumpp}},
  \bibinfo {author} {\bibfnamefont {A.~A.}\ \bibnamefont {Sorokin}}, \bibinfo
  {author} {\bibfnamefont {K.}~\bibnamefont {Tiedtke}}, \bibinfo {author}
  {\bibfnamefont {M.}~\bibnamefont {Richter}}, \bibinfo {author} {\bibfnamefont
  {V.}~\bibnamefont {B{\"{u}}rk}}, \bibinfo {author} {\bibfnamefont
  {K.}~\bibnamefont {Mertens}}, \bibinfo {author} {\bibfnamefont
  {P.}~\bibnamefont {Jurani\'{c}}}, \ and\ \bibinfo {author} {\bibfnamefont
  {M.}~\bibnamefont {Martins}},\ }\href {\doibase
  10.1103/PhysRevLett.112.213002} {\bibfield  {journal} {\bibinfo  {journal}
  {Phys. Rev. Lett.}\ }\textbf {\bibinfo {volume} {112}},\ \bibinfo {pages}
  {213002} (\bibinfo {year} {2014})}\BibitemShut {NoStop}%
\bibitem [{\citenamefont {Rudenko}\ \emph {et~al.}(2017)\citenamefont
  {Rudenko}, \citenamefont {Inhester}, \citenamefont {Hanasaki}, \citenamefont
  {Li}, \citenamefont {Robatjazi}, \citenamefont {Erk}, \citenamefont {Boll},
  \citenamefont {Toyota}, \citenamefont {Hao}, \citenamefont {Vendrell},
  \citenamefont {Bomme}, \citenamefont {Savelyev}, \citenamefont {Rudek},
  \citenamefont {Foucar}, \citenamefont {Southworth}, \citenamefont {Lehmann},
  \citenamefont {Kraessig}, \citenamefont {Marchenko}, \citenamefont {Simon},
  \citenamefont {Ueda}, \citenamefont {Ferguson}, \citenamefont {Bucher},
  \citenamefont {Gorkhover}, \citenamefont {Carron}, \citenamefont
  {Alonso-Mori}, \citenamefont {Koglin}, \citenamefont {Correa}, \citenamefont
  {Williams}, \citenamefont {Boutet}, \citenamefont {Young}, \citenamefont
  {Bostedt}, \citenamefont {Son}, \citenamefont {Santra},\ and\ \citenamefont
  {Rolles}}]{Rudenko2017}%
  \BibitemOpen
  \bibfield  {author} {\bibinfo {author} {\bibfnamefont {A.}~\bibnamefont
  {Rudenko}}, \bibinfo {author} {\bibfnamefont {L.}~\bibnamefont {Inhester}},
  \bibinfo {author} {\bibfnamefont {K.}~\bibnamefont {Hanasaki}}, \bibinfo
  {author} {\bibfnamefont {X.}~\bibnamefont {Li}}, \bibinfo {author}
  {\bibfnamefont {S.~J.}\ \bibnamefont {Robatjazi}}, \bibinfo {author}
  {\bibfnamefont {B.}~\bibnamefont {Erk}}, \bibinfo {author} {\bibfnamefont
  {R.}~\bibnamefont {Boll}}, \bibinfo {author} {\bibfnamefont {K.}~\bibnamefont
  {Toyota}}, \bibinfo {author} {\bibfnamefont {Y.}~\bibnamefont {Hao}},
  \bibinfo {author} {\bibfnamefont {O.}~\bibnamefont {Vendrell}}, \bibinfo
  {author} {\bibfnamefont {C.}~\bibnamefont {Bomme}}, \bibinfo {author}
  {\bibfnamefont {E.}~\bibnamefont {Savelyev}}, \bibinfo {author}
  {\bibfnamefont {B.}~\bibnamefont {Rudek}}, \bibinfo {author} {\bibfnamefont
  {L.}~\bibnamefont {Foucar}}, \bibinfo {author} {\bibfnamefont {S.~H.}\
  \bibnamefont {Southworth}}, \bibinfo {author} {\bibfnamefont {C.~S.}\
  \bibnamefont {Lehmann}}, \bibinfo {author} {\bibfnamefont {B.}~\bibnamefont
  {Kraessig}}, \bibinfo {author} {\bibfnamefont {T.}~\bibnamefont {Marchenko}},
  \bibinfo {author} {\bibfnamefont {M.}~\bibnamefont {Simon}}, \bibinfo
  {author} {\bibfnamefont {K.}~\bibnamefont {Ueda}}, \bibinfo {author}
  {\bibfnamefont {K.~R.}\ \bibnamefont {Ferguson}}, \bibinfo {author}
  {\bibfnamefont {M.}~\bibnamefont {Bucher}}, \bibinfo {author} {\bibfnamefont
  {T.}~\bibnamefont {Gorkhover}}, \bibinfo {author} {\bibfnamefont
  {S.}~\bibnamefont {Carron}}, \bibinfo {author} {\bibfnamefont
  {R.}~\bibnamefont {Alonso-Mori}}, \bibinfo {author} {\bibfnamefont {J.~E.}\
  \bibnamefont {Koglin}}, \bibinfo {author} {\bibfnamefont {J.}~\bibnamefont
  {Correa}}, \bibinfo {author} {\bibfnamefont {G.~J.}\ \bibnamefont
  {Williams}}, \bibinfo {author} {\bibfnamefont {S.}~\bibnamefont {Boutet}},
  \bibinfo {author} {\bibfnamefont {L.}~\bibnamefont {Young}}, \bibinfo
  {author} {\bibfnamefont {C.}~\bibnamefont {Bostedt}}, \bibinfo {author}
  {\bibfnamefont {S.-K.}\ \bibnamefont {Son}}, \bibinfo {author} {\bibfnamefont
  {R.}~\bibnamefont {Santra}}, \ and\ \bibinfo {author} {\bibfnamefont
  {D.}~\bibnamefont {Rolles}},\ }\href {\doibase 10.1038/nature22373}
  {\bibfield  {journal} {\bibinfo  {journal} {Nature}\ }\textbf {\bibinfo
  {volume} {546}},\ \bibinfo {pages} {129} (\bibinfo {year}
  {2017})}\BibitemShut {NoStop}%
\bibitem [{\citenamefont {Tamasaku}\ \emph {et~al.}(2013)\citenamefont
  {Tamasaku}, \citenamefont {Nagasono}, \citenamefont {Iwayama}, \citenamefont
  {Shigemasa}, \citenamefont {Inubushi}, \citenamefont {Tanaka}, \citenamefont
  {Tono}, \citenamefont {Togashi}, \citenamefont {Sato}, \citenamefont
  {Katayama}, \citenamefont {Kameshima}, \citenamefont {Hatsui}, \citenamefont
  {Yabashi},\ and\ \citenamefont {Ishikawa}}]{Tamasaku2013}%
  \BibitemOpen
  \bibfield  {author} {\bibinfo {author} {\bibfnamefont {K.}~\bibnamefont
  {Tamasaku}}, \bibinfo {author} {\bibfnamefont {M.}~\bibnamefont {Nagasono}},
  \bibinfo {author} {\bibfnamefont {H.}~\bibnamefont {Iwayama}}, \bibinfo
  {author} {\bibfnamefont {E.}~\bibnamefont {Shigemasa}}, \bibinfo {author}
  {\bibfnamefont {Y.}~\bibnamefont {Inubushi}}, \bibinfo {author}
  {\bibfnamefont {T.}~\bibnamefont {Tanaka}}, \bibinfo {author} {\bibfnamefont
  {K.}~\bibnamefont {Tono}}, \bibinfo {author} {\bibfnamefont {T.}~\bibnamefont
  {Togashi}}, \bibinfo {author} {\bibfnamefont {T.}~\bibnamefont {Sato}},
  \bibinfo {author} {\bibfnamefont {T.}~\bibnamefont {Katayama}}, \bibinfo
  {author} {\bibfnamefont {T.}~\bibnamefont {Kameshima}}, \bibinfo {author}
  {\bibfnamefont {T.}~\bibnamefont {Hatsui}}, \bibinfo {author} {\bibfnamefont
  {M.}~\bibnamefont {Yabashi}}, \ and\ \bibinfo {author} {\bibfnamefont
  {T.}~\bibnamefont {Ishikawa}},\ }\href {\doibase
  10.1103/PhysRevLett.111.043001} {\bibfield  {journal} {\bibinfo  {journal}
  {Physical Review Letters}\ }\textbf {\bibinfo {volume} {111}},\ \bibinfo
  {pages} {43001} (\bibinfo {year} {2013})}\BibitemShut {NoStop}%
\bibitem [{\citenamefont {Fang}\ \emph {et~al.}(2010)\citenamefont {Fang},
  \citenamefont {Hoener}, \citenamefont {Gessner}, \citenamefont {Tarantelli},
  \citenamefont {Pratt}, \citenamefont {Kornilov}, \citenamefont {Buth},
  \citenamefont {G{\"{u}}hr}, \citenamefont {Kanter}, \citenamefont {Bostedt},
  \citenamefont {Bozek}, \citenamefont {Bucksbaum}, \citenamefont {Chen},
  \citenamefont {Coffee}, \citenamefont {Cryan}, \citenamefont {Glownia},
  \citenamefont {Kukk}, \citenamefont {Leone},\ and\ \citenamefont
  {Berrah}}]{Fang2010}%
  \BibitemOpen
  \bibfield  {author} {\bibinfo {author} {\bibfnamefont {L.}~\bibnamefont
  {Fang}}, \bibinfo {author} {\bibfnamefont {M.}~\bibnamefont {Hoener}},
  \bibinfo {author} {\bibfnamefont {O.}~\bibnamefont {Gessner}}, \bibinfo
  {author} {\bibfnamefont {F.}~\bibnamefont {Tarantelli}}, \bibinfo {author}
  {\bibfnamefont {S.~T.}\ \bibnamefont {Pratt}}, \bibinfo {author}
  {\bibfnamefont {O.}~\bibnamefont {Kornilov}}, \bibinfo {author}
  {\bibfnamefont {C.}~\bibnamefont {Buth}}, \bibinfo {author} {\bibfnamefont
  {M.}~\bibnamefont {G{\"{u}}hr}}, \bibinfo {author} {\bibfnamefont {E.~P.}\
  \bibnamefont {Kanter}}, \bibinfo {author} {\bibfnamefont {C.}~\bibnamefont
  {Bostedt}}, \bibinfo {author} {\bibfnamefont {J.~D.}\ \bibnamefont {Bozek}},
  \bibinfo {author} {\bibfnamefont {P.~H.}\ \bibnamefont {Bucksbaum}}, \bibinfo
  {author} {\bibfnamefont {M.}~\bibnamefont {Chen}}, \bibinfo {author}
  {\bibfnamefont {R.}~\bibnamefont {Coffee}}, \bibinfo {author} {\bibfnamefont
  {J.}~\bibnamefont {Cryan}}, \bibinfo {author} {\bibfnamefont
  {M.}~\bibnamefont {Glownia}}, \bibinfo {author} {\bibfnamefont
  {E.}~\bibnamefont {Kukk}}, \bibinfo {author} {\bibfnamefont {S.~R.}\
  \bibnamefont {Leone}}, \ and\ \bibinfo {author} {\bibfnamefont
  {N.}~\bibnamefont {Berrah}},\ }\href {\doibase
  10.1103/PhysRevLett.105.083005} {\bibfield  {journal} {\bibinfo  {journal}
  {Physical Review Letters}\ }\textbf {\bibinfo {volume} {105}},\ \bibinfo
  {pages} {83005} (\bibinfo {year} {2010})}\BibitemShut {NoStop}%
\bibitem [{\citenamefont {Doumy}\ \emph {et~al.}(2011)\citenamefont {Doumy},
  \citenamefont {Roedig}, \citenamefont {Son}, \citenamefont {Blaga},
  \citenamefont {DiChiara}, \citenamefont {Santra}, \citenamefont {Berrah},
  \citenamefont {Bostedt}, \citenamefont {Bozek}, \citenamefont {Bucksbaum},
  \citenamefont {Cryan}, \citenamefont {Fang}, \citenamefont {Ghimire},
  \citenamefont {Glownia}, \citenamefont {Hoener}, \citenamefont {Kanter},
  \citenamefont {Kr{\"{a}}ssig}, \citenamefont {Kuebel}, \citenamefont
  {Messerschmidt}, \citenamefont {Paulus}, \citenamefont {Reis}, \citenamefont
  {Rohringer}, \citenamefont {Young}, \citenamefont {Agostini},\ and\
  \citenamefont {DiMauro}}]{ghimi}%
  \BibitemOpen
  \bibfield  {author} {\bibinfo {author} {\bibfnamefont {G.}~\bibnamefont
  {Doumy}}, \bibinfo {author} {\bibfnamefont {C.}~\bibnamefont {Roedig}},
  \bibinfo {author} {\bibfnamefont {S.-K.}\ \bibnamefont {Son}}, \bibinfo
  {author} {\bibfnamefont {C.~I.}\ \bibnamefont {Blaga}}, \bibinfo {author}
  {\bibfnamefont {A.~D.}\ \bibnamefont {DiChiara}}, \bibinfo {author}
  {\bibfnamefont {R.}~\bibnamefont {Santra}}, \bibinfo {author} {\bibfnamefont
  {N.}~\bibnamefont {Berrah}}, \bibinfo {author} {\bibfnamefont
  {C.}~\bibnamefont {Bostedt}}, \bibinfo {author} {\bibfnamefont {J.~D.}\
  \bibnamefont {Bozek}}, \bibinfo {author} {\bibfnamefont {P.~H.}\ \bibnamefont
  {Bucksbaum}}, \bibinfo {author} {\bibfnamefont {J.~P.}\ \bibnamefont
  {Cryan}}, \bibinfo {author} {\bibfnamefont {L.}~\bibnamefont {Fang}},
  \bibinfo {author} {\bibfnamefont {S.}~\bibnamefont {Ghimire}}, \bibinfo
  {author} {\bibfnamefont {J.~M.}\ \bibnamefont {Glownia}}, \bibinfo {author}
  {\bibfnamefont {M.}~\bibnamefont {Hoener}}, \bibinfo {author} {\bibfnamefont
  {E.~P.}\ \bibnamefont {Kanter}}, \bibinfo {author} {\bibfnamefont
  {B.}~\bibnamefont {Kr{\"{a}}ssig}}, \bibinfo {author} {\bibfnamefont
  {M.}~\bibnamefont {Kuebel}}, \bibinfo {author} {\bibfnamefont
  {M.}~\bibnamefont {Messerschmidt}}, \bibinfo {author} {\bibfnamefont {G.~G.}\
  \bibnamefont {Paulus}}, \bibinfo {author} {\bibfnamefont {D.~A.}\
  \bibnamefont {Reis}}, \bibinfo {author} {\bibfnamefont {N.}~\bibnamefont
  {Rohringer}}, \bibinfo {author} {\bibfnamefont {L.}~\bibnamefont {Young}},
  \bibinfo {author} {\bibfnamefont {P.}~\bibnamefont {Agostini}}, \ and\
  \bibinfo {author} {\bibfnamefont {L.~F.}\ \bibnamefont {DiMauro}},\ }\href
  {\doibase 10.1103/PhysRevLett.106.083002} {\bibfield  {journal} {\bibinfo
  {journal} {Phys. Rev. Lett.}\ }\textbf {\bibinfo {volume} {106}},\ \bibinfo
  {pages} {83002} (\bibinfo {year} {2011})}\BibitemShut {NoStop}%
\bibitem [{\citenamefont {Tamasaku}\ \emph {et~al.}(2014)\citenamefont
  {Tamasaku}, \citenamefont {Shigemasa}, \citenamefont {Inubushi},
  \citenamefont {Katayama}, \citenamefont {Sawada}, \citenamefont {Yumoto},
  \citenamefont {Ohashi}, \citenamefont {Mimura}, \citenamefont {Yabashi},
  \citenamefont {Yamauchi},\ and\ \citenamefont {Ishikawa}}]{Tamasaku2014}%
  \BibitemOpen
  \bibfield  {author} {\bibinfo {author} {\bibfnamefont {K.}~\bibnamefont
  {Tamasaku}}, \bibinfo {author} {\bibfnamefont {E.}~\bibnamefont {Shigemasa}},
  \bibinfo {author} {\bibfnamefont {Y.}~\bibnamefont {Inubushi}}, \bibinfo
  {author} {\bibfnamefont {T.}~\bibnamefont {Katayama}}, \bibinfo {author}
  {\bibfnamefont {K.}~\bibnamefont {Sawada}}, \bibinfo {author} {\bibfnamefont
  {H.}~\bibnamefont {Yumoto}}, \bibinfo {author} {\bibfnamefont
  {H.}~\bibnamefont {Ohashi}}, \bibinfo {author} {\bibfnamefont
  {H.}~\bibnamefont {Mimura}}, \bibinfo {author} {\bibfnamefont
  {M.}~\bibnamefont {Yabashi}}, \bibinfo {author} {\bibfnamefont
  {K.}~\bibnamefont {Yamauchi}}, \ and\ \bibinfo {author} {\bibfnamefont
  {T.}~\bibnamefont {Ishikawa}},\ }\href {\doibase 10.1038/nphoton.2014.10}
  {\bibfield  {journal} {\bibinfo  {journal} {Nature Photonics}\ }\textbf
  {\bibinfo {volume} {8}},\ \bibinfo {pages} {313} (\bibinfo {year}
  {2014})}\BibitemShut {NoStop}%
\bibitem [{\citenamefont {Ghimire}\ \emph {et~al.}(2016)\citenamefont
  {Ghimire}, \citenamefont {Fuchs}, \citenamefont {Hastings}, \citenamefont
  {Herrmann}, \citenamefont {Inubushi}, \citenamefont {Pines}, \citenamefont
  {Shwartz}, \citenamefont {Yabashi},\ and\ \citenamefont
  {Reis}}]{Ghimire2016}%
  \BibitemOpen
  \bibfield  {author} {\bibinfo {author} {\bibfnamefont {S.}~\bibnamefont
  {Ghimire}}, \bibinfo {author} {\bibfnamefont {M.}~\bibnamefont {Fuchs}},
  \bibinfo {author} {\bibfnamefont {J.}~\bibnamefont {Hastings}}, \bibinfo
  {author} {\bibfnamefont {S.~C.}\ \bibnamefont {Herrmann}}, \bibinfo {author}
  {\bibfnamefont {Y.}~\bibnamefont {Inubushi}}, \bibinfo {author}
  {\bibfnamefont {J.}~\bibnamefont {Pines}}, \bibinfo {author} {\bibfnamefont
  {S.}~\bibnamefont {Shwartz}}, \bibinfo {author} {\bibfnamefont
  {M.}~\bibnamefont {Yabashi}}, \ and\ \bibinfo {author} {\bibfnamefont
  {D.~A.}\ \bibnamefont {Reis}},\ }\href {\doibase 10.1103/PhysRevA.94.043418}
  {\bibfield  {journal} {\bibinfo  {journal} {Physical Review A}\ }\textbf
  {\bibinfo {volume} {94}},\ \bibinfo {pages} {43418} (\bibinfo {year}
  {2016})}\BibitemShut {NoStop}%
\bibitem [{\citenamefont {Weninger}\ \emph {et~al.}(2013)\citenamefont
  {Weninger}, \citenamefont {Purvis}, \citenamefont {Ryan}, \citenamefont
  {London}, \citenamefont {Bozek}, \citenamefont {Bostedt}, \citenamefont
  {Graf}, \citenamefont {Brown}, \citenamefont {Rocca},\ and\ \citenamefont
  {Rohringer}}]{Weninger2013}%
  \BibitemOpen
  \bibfield  {author} {\bibinfo {author} {\bibfnamefont {C.}~\bibnamefont
  {Weninger}}, \bibinfo {author} {\bibfnamefont {M.}~\bibnamefont {Purvis}},
  \bibinfo {author} {\bibfnamefont {D.}~\bibnamefont {Ryan}}, \bibinfo {author}
  {\bibfnamefont {R.~A.}\ \bibnamefont {London}}, \bibinfo {author}
  {\bibfnamefont {J.~D.}\ \bibnamefont {Bozek}}, \bibinfo {author}
  {\bibfnamefont {C.}~\bibnamefont {Bostedt}}, \bibinfo {author} {\bibfnamefont
  {A.}~\bibnamefont {Graf}}, \bibinfo {author} {\bibfnamefont {G.}~\bibnamefont
  {Brown}}, \bibinfo {author} {\bibfnamefont {J.~J.}\ \bibnamefont {Rocca}}, \
  and\ \bibinfo {author} {\bibfnamefont {N.}~\bibnamefont {Rohringer}},\ }\href
  {\doibase 10.1103/PhysRevLett.111.233902} {\bibfield  {journal} {\bibinfo
  {journal} {Physical Review Letters}\ }\textbf {\bibinfo {volume} {111}},\
  \bibinfo {pages} {233902} (\bibinfo {year} {2013})}\BibitemShut {NoStop}%
\bibitem [{\citenamefont {H.}\ \emph {et~al.}(2015)\citenamefont {H.},
  \citenamefont {Inubushi}, \citenamefont {Nagamine}, \citenamefont {Michine},
  \citenamefont {Ohashi}, \citenamefont {{Yumoto H. Yamauchi}}, \citenamefont
  {Mimura}, \citenamefont {Kitamura}, \citenamefont {Katayama}, \citenamefont
  {Ishikawa},\ and\ \citenamefont {Yabashi}}]{Yoneda}%
  \BibitemOpen
  \bibfield  {author} {\bibinfo {author} {\bibfnamefont {Y.}~\bibnamefont
  {H.}}, \bibinfo {author} {\bibfnamefont {Y.}~\bibnamefont {Inubushi}},
  \bibinfo {author} {\bibfnamefont {K.}~\bibnamefont {Nagamine}}, \bibinfo
  {author} {\bibfnamefont {Y.}~\bibnamefont {Michine}}, \bibinfo {author}
  {\bibfnamefont {H.}~\bibnamefont {Ohashi}}, \bibinfo {author} {\bibfnamefont
  {K.}~\bibnamefont {{Yumoto H. Yamauchi}}}, \bibinfo {author} {\bibfnamefont
  {H.}~\bibnamefont {Mimura}}, \bibinfo {author} {\bibfnamefont
  {H.}~\bibnamefont {Kitamura}}, \bibinfo {author} {\bibfnamefont
  {T.}~\bibnamefont {Katayama}}, \bibinfo {author} {\bibfnamefont
  {T.}~\bibnamefont {Ishikawa}}, \ and\ \bibinfo {author} {\bibfnamefont
  {M.}~\bibnamefont {Yabashi}},\ }\href {\doibase 10.1038/nature14894}
  {\bibfield  {journal} {\bibinfo  {journal} {Nature}\ }\textbf {\bibinfo
  {volume} {524}},\ \bibinfo {pages} {446} (\bibinfo {year}
  {2015})}\BibitemShut {NoStop}%
\bibitem [{\citenamefont {Kroll}\ \emph {et~al.}(2018)\citenamefont {Kroll},
  \citenamefont {Weninger}, \citenamefont {Alonso-Mori}, \citenamefont
  {Sokaras}, \citenamefont {Zhu}, \citenamefont {Mercadier}, \citenamefont
  {Majety}, \citenamefont {Marinelli}, \citenamefont {Lutman}, \citenamefont
  {Guetg}, \citenamefont {Decker}, \citenamefont {Boutet}, \citenamefont
  {Aquila}, \citenamefont {Koglin}, \citenamefont {Koralek}, \citenamefont
  {DePonte}, \citenamefont {Kern}, \citenamefont {Fuller}, \citenamefont
  {Pastor}, \citenamefont {Fransson}, \citenamefont {Zhang}, \citenamefont
  {Yano}, \citenamefont {Yachandra}, \citenamefont {Rohringer},\ and\
  \citenamefont {Bergmann}}]{kroll}%
  \BibitemOpen
  \bibfield  {author} {\bibinfo {author} {\bibfnamefont {T.}~\bibnamefont
  {Kroll}}, \bibinfo {author} {\bibfnamefont {C.}~\bibnamefont {Weninger}},
  \bibinfo {author} {\bibfnamefont {R.}~\bibnamefont {Alonso-Mori}}, \bibinfo
  {author} {\bibfnamefont {D.}~\bibnamefont {Sokaras}}, \bibinfo {author}
  {\bibfnamefont {D.}~\bibnamefont {Zhu}}, \bibinfo {author} {\bibfnamefont
  {L.}~\bibnamefont {Mercadier}}, \bibinfo {author} {\bibfnamefont {V.~P.}\
  \bibnamefont {Majety}}, \bibinfo {author} {\bibfnamefont {A.}~\bibnamefont
  {Marinelli}}, \bibinfo {author} {\bibfnamefont {A.}~\bibnamefont {Lutman}},
  \bibinfo {author} {\bibfnamefont {M.~W.}\ \bibnamefont {Guetg}}, \bibinfo
  {author} {\bibfnamefont {F.-J.}\ \bibnamefont {Decker}}, \bibinfo {author}
  {\bibfnamefont {S.}~\bibnamefont {Boutet}}, \bibinfo {author} {\bibfnamefont
  {A.}~\bibnamefont {Aquila}}, \bibinfo {author} {\bibfnamefont
  {J.}~\bibnamefont {Koglin}}, \bibinfo {author} {\bibfnamefont
  {J.}~\bibnamefont {Koralek}}, \bibinfo {author} {\bibfnamefont {D.~P.}\
  \bibnamefont {DePonte}}, \bibinfo {author} {\bibfnamefont {J.}~\bibnamefont
  {Kern}}, \bibinfo {author} {\bibfnamefont {F.~D.}\ \bibnamefont {Fuller}},
  \bibinfo {author} {\bibfnamefont {E.}~\bibnamefont {Pastor}}, \bibinfo
  {author} {\bibfnamefont {T.}~\bibnamefont {Fransson}}, \bibinfo {author}
  {\bibfnamefont {Y.}~\bibnamefont {Zhang}}, \bibinfo {author} {\bibfnamefont
  {J.}~\bibnamefont {Yano}}, \bibinfo {author} {\bibfnamefont {V.~K.}\
  \bibnamefont {Yachandra}}, \bibinfo {author} {\bibfnamefont {N.}~\bibnamefont
  {Rohringer}}, \ and\ \bibinfo {author} {\bibfnamefont {U.}~\bibnamefont
  {Bergmann}},\ }\href {\doibase 10.1103/PhysRevLett.120.133203} {\bibfield
  {journal} {\bibinfo  {journal} {Phys. Rev. Lett.}\ }\textbf {\bibinfo
  {volume} {120}},\ \bibinfo {pages} {133203} (\bibinfo {year}
  {2018})}\BibitemShut {NoStop}%
\bibitem [{\citenamefont {Rohringer}\ \emph {et~al.}(2012)\citenamefont
  {Rohringer}, \citenamefont {Duncan}, \citenamefont {London}, \citenamefont
  {Purvis}, \citenamefont {Albert}, \citenamefont {Dunn}, \citenamefont
  {Bozek}, \citenamefont {Bostedt}, \citenamefont {Graf}, \citenamefont {Hill},
  \citenamefont {S.P.},\ and\ \citenamefont {Rocca}}]{rohringer}%
  \BibitemOpen
  \bibfield  {author} {\bibinfo {author} {\bibfnamefont {N.}~\bibnamefont
  {Rohringer}}, \bibinfo {author} {\bibfnamefont {R.}~\bibnamefont {Duncan}},
  \bibinfo {author} {\bibfnamefont {R.~A.}\ \bibnamefont {London}}, \bibinfo
  {author} {\bibfnamefont {M.}~\bibnamefont {Purvis}}, \bibinfo {author}
  {\bibfnamefont {F.}~\bibnamefont {Albert}}, \bibinfo {author} {\bibfnamefont
  {J.}~\bibnamefont {Dunn}}, \bibinfo {author} {\bibfnamefont {J.~D.}\
  \bibnamefont {Bozek}}, \bibinfo {author} {\bibfnamefont {C.}~\bibnamefont
  {Bostedt}}, \bibinfo {author} {\bibfnamefont {A.}~\bibnamefont {Graf}},
  \bibinfo {author} {\bibfnamefont {R.}~\bibnamefont {Hill}}, \bibinfo {author}
  {\bibfnamefont {H.-R.}\ \bibnamefont {S.P.}}, \ and\ \bibinfo {author}
  {\bibfnamefont {J.~J.}\ \bibnamefont {Rocca}},\ }\href {\doibase
  10.1038/nature10721} {\bibfield  {journal} {\bibinfo  {journal} {Nature}\
  }\textbf {\bibinfo {volume} {481}},\ \bibinfo {pages} {488} (\bibinfo {year}
  {2012})}\BibitemShut {NoStop}%
\bibitem [{\citenamefont {Shwartz}\ \emph {et~al.}(2014)\citenamefont
  {Shwartz}, \citenamefont {Fuchs}, \citenamefont {Hastings}, \citenamefont
  {Inubushi}, \citenamefont {Ishikawa}, \citenamefont {Katayama}, \citenamefont
  {Reis}, \citenamefont {Sato}, \citenamefont {Tono}, \citenamefont {Yabashi},
  \citenamefont {Yudovich},\ and\ \citenamefont {Harris}}]{shwartz1}%
  \BibitemOpen
  \bibfield  {author} {\bibinfo {author} {\bibfnamefont {S.}~\bibnamefont
  {Shwartz}}, \bibinfo {author} {\bibfnamefont {M.}~\bibnamefont {Fuchs}},
  \bibinfo {author} {\bibfnamefont {J.~B.}\ \bibnamefont {Hastings}}, \bibinfo
  {author} {\bibfnamefont {Y.}~\bibnamefont {Inubushi}}, \bibinfo {author}
  {\bibfnamefont {T.}~\bibnamefont {Ishikawa}}, \bibinfo {author}
  {\bibfnamefont {T.}~\bibnamefont {Katayama}}, \bibinfo {author}
  {\bibfnamefont {D.~A.}\ \bibnamefont {Reis}}, \bibinfo {author}
  {\bibfnamefont {T.}~\bibnamefont {Sato}}, \bibinfo {author} {\bibfnamefont
  {K.}~\bibnamefont {Tono}}, \bibinfo {author} {\bibfnamefont {M.}~\bibnamefont
  {Yabashi}}, \bibinfo {author} {\bibfnamefont {S.}~\bibnamefont {Yudovich}}, \
  and\ \bibinfo {author} {\bibfnamefont {S.~E.}\ \bibnamefont {Harris}},\
  }\href {\doibase 10.1103/PhysRevLett.112.163901} {\bibfield  {journal}
  {\bibinfo  {journal} {Phys. Rev. Lett.}\ }\textbf {\bibinfo {volume} {112}},\
  \bibinfo {pages} {163901} (\bibinfo {year} {2014})}\BibitemShut {NoStop}%
\bibitem [{\citenamefont {Fuchs}\ \emph {et~al.}(2015)\citenamefont {Fuchs},
  \citenamefont {Trigo}, \citenamefont {Chen}, \citenamefont {Ghimire},
  \citenamefont {Shwartz}, \citenamefont {Kizina}, \citenamefont {Jiang},
  \citenamefont {Henighan}, \citenamefont {Bray}, \citenamefont {Ndabashimiye},
  \citenamefont {Bucksbaum}, \citenamefont {Feng}, \citenamefont {Herrmann},
  \citenamefont {Carini}, \citenamefont {Pines}, \citenamefont {Hart},
  \citenamefont {Kenney}, \citenamefont {Guillet}, \citenamefont {Boutet},
  \citenamefont {Williams}, \citenamefont {Messerschmidt}, \citenamefont
  {Seibert}, \citenamefont {Moeller}, \citenamefont {Hastings},\ and\
  \citenamefont {Reis}}]{compton}%
  \BibitemOpen
  \bibfield  {author} {\bibinfo {author} {\bibfnamefont {M.}~\bibnamefont
  {Fuchs}}, \bibinfo {author} {\bibfnamefont {M.}~\bibnamefont {Trigo}},
  \bibinfo {author} {\bibfnamefont {J.}~\bibnamefont {Chen}}, \bibinfo {author}
  {\bibfnamefont {S.}~\bibnamefont {Ghimire}}, \bibinfo {author} {\bibfnamefont
  {S.}~\bibnamefont {Shwartz}}, \bibinfo {author} {\bibfnamefont
  {M.}~\bibnamefont {Kizina}}, \bibinfo {author} {\bibfnamefont
  {M.}~\bibnamefont {Jiang}}, \bibinfo {author} {\bibfnamefont
  {T.}~\bibnamefont {Henighan}}, \bibinfo {author} {\bibfnamefont
  {C.}~\bibnamefont {Bray}}, \bibinfo {author} {\bibfnamefont {G.}~\bibnamefont
  {Ndabashimiye}}, \bibinfo {author} {\bibfnamefont {P.~H.}\ \bibnamefont
  {Bucksbaum}}, \bibinfo {author} {\bibfnamefont {Y.}~\bibnamefont {Feng}},
  \bibinfo {author} {\bibfnamefont {S.}~\bibnamefont {Herrmann}}, \bibinfo
  {author} {\bibfnamefont {G.~A.}\ \bibnamefont {Carini}}, \bibinfo {author}
  {\bibfnamefont {J.}~\bibnamefont {Pines}}, \bibinfo {author} {\bibfnamefont
  {P.}~\bibnamefont {Hart}}, \bibinfo {author} {\bibfnamefont {C.}~\bibnamefont
  {Kenney}}, \bibinfo {author} {\bibfnamefont {S.}~\bibnamefont {Guillet}},
  \bibinfo {author} {\bibfnamefont {S.}~\bibnamefont {Boutet}}, \bibinfo
  {author} {\bibfnamefont {G.}~\bibnamefont {Williams}}, \bibinfo {author}
  {\bibfnamefont {M.}~\bibnamefont {Messerschmidt}}, \bibinfo {author}
  {\bibfnamefont {M.~M.}\ \bibnamefont {Seibert}}, \bibinfo {author}
  {\bibfnamefont {S.}~\bibnamefont {Moeller}}, \bibinfo {author} {\bibfnamefont
  {J.~B.}\ \bibnamefont {Hastings}}, \ and\ \bibinfo {author} {\bibfnamefont
  {D.~A.}\ \bibnamefont {Reis}},\ }\href {\doibase 10.1038/nphys3452}
  {\bibfield  {journal} {\bibinfo  {journal} {Nature Physics}\ }\textbf
  {\bibinfo {volume} {11}},\ \bibinfo {pages} {964} (\bibinfo {year}
  {2015})}\BibitemShut {NoStop}%
\bibitem [{\citenamefont {Krebs}\ \emph {et~al.}(2019)\citenamefont {Krebs},
  \citenamefont {Reis},\ and\ \citenamefont {Santra}}]{Krebs2019}%
  \BibitemOpen
  \bibfield  {author} {\bibinfo {author} {\bibfnamefont {D.}~\bibnamefont
  {Krebs}}, \bibinfo {author} {\bibfnamefont {D.~A.}\ \bibnamefont {Reis}}, \
  and\ \bibinfo {author} {\bibfnamefont {R.}~\bibnamefont {Santra}},\ }\href
  {\doibase 10.1103/PhysRevA.99.022120} {\bibfield  {journal} {\bibinfo
  {journal} {Phys. Rev. A}\ }\textbf {\bibinfo {volume} {99}},\ \bibinfo
  {pages} {22120} (\bibinfo {year} {2019})}\BibitemShut {NoStop}%
\bibitem [{\citenamefont {Venkatesh}\ and\ \citenamefont
  {Robicheaux}(2020)}]{Venkatesh2020}%
  \BibitemOpen
  \bibfield  {author} {\bibinfo {author} {\bibfnamefont {A.}~\bibnamefont
  {Venkatesh}}\ and\ \bibinfo {author} {\bibfnamefont {F.}~\bibnamefont
  {Robicheaux}},\ }\href {\doibase 10.1103/PhysRevA.101.013409} {\bibfield
  {journal} {\bibinfo  {journal} {Phys. Rev. A}\ }\textbf {\bibinfo {volume}
  {101}},\ \bibinfo {pages} {13409} (\bibinfo {year} {2020})}\BibitemShut
  {NoStop}%
\bibitem [{\citenamefont {Hoszowska}\ \emph {et~al.}(2011)\citenamefont
  {Hoszowska}, \citenamefont {Dousse}, \citenamefont {Szlachetko},
  \citenamefont {Kayser}, \citenamefont {Cao}, \citenamefont
  {Jagodzi{\'{n}}ski}, \citenamefont {Kav{\v{c}}i{\v{c}}},\ and\ \citenamefont
  {Nowak}}]{Hoszowska2011}%
  \BibitemOpen
  \bibfield  {author} {\bibinfo {author} {\bibfnamefont {J.}~\bibnamefont
  {Hoszowska}}, \bibinfo {author} {\bibfnamefont {J.-C.}\ \bibnamefont
  {Dousse}}, \bibinfo {author} {\bibfnamefont {J.}~\bibnamefont {Szlachetko}},
  \bibinfo {author} {\bibfnamefont {Y.}~\bibnamefont {Kayser}}, \bibinfo
  {author} {\bibfnamefont {W.}~\bibnamefont {Cao}}, \bibinfo {author}
  {\bibfnamefont {P.}~\bibnamefont {Jagodzi{\'{n}}ski}}, \bibinfo {author}
  {\bibfnamefont {M.}~\bibnamefont {Kav{\v{c}}i{\v{c}}}}, \ and\ \bibinfo
  {author} {\bibfnamefont {S.~H.}\ \bibnamefont {Nowak}},\ }\href {\doibase
  10.1103/PhysRevLett.107.053001} {\bibfield  {journal} {\bibinfo  {journal}
  {Physical Review Letters}\ }\textbf {\bibinfo {volume} {107}},\ \bibinfo
  {pages} {53001} (\bibinfo {year} {2011})}\BibitemShut {NoStop}%
\bibitem [{\citenamefont {Terhune}\ \emph {et~al.}(1965)\citenamefont
  {Terhune}, \citenamefont {Maker},\ and\ \citenamefont
  {Savage}}]{Terhune1965}%
  \BibitemOpen
  \bibfield  {author} {\bibinfo {author} {\bibfnamefont {R.~W.}\ \bibnamefont
  {Terhune}}, \bibinfo {author} {\bibfnamefont {P.~D.}\ \bibnamefont {Maker}},
  \ and\ \bibinfo {author} {\bibfnamefont {C.~M.}\ \bibnamefont {Savage}},\
  }\href {\doibase 10.1103/PhysRevLett.14.681} {\bibfield  {journal} {\bibinfo
  {journal} {Phys. Rev. Lett.}\ }\textbf {\bibinfo {volume} {14}},\ \bibinfo
  {pages} {681} (\bibinfo {year} {1965})}\BibitemShut {NoStop}%
\bibitem [{\citenamefont {Tono}\ \emph {et~al.}(2013)\citenamefont {Tono},
  \citenamefont {Togashi}, \citenamefont {Inubushi}, \citenamefont {Sato},
  \citenamefont {Katayama}, \citenamefont {Ogawa}, \citenamefont {Ohashi},
  \citenamefont {Kimura}, \citenamefont {Takahashi}, \citenamefont {Takeshita},
  \citenamefont {Tomizawa}, \citenamefont {Goto}, \citenamefont {Ishikawa},\
  and\ \citenamefont {Yabashi}}]{Tono2013}%
  \BibitemOpen
  \bibfield  {author} {\bibinfo {author} {\bibfnamefont {K.}~\bibnamefont
  {Tono}}, \bibinfo {author} {\bibfnamefont {T.}~\bibnamefont {Togashi}},
  \bibinfo {author} {\bibfnamefont {Y.}~\bibnamefont {Inubushi}}, \bibinfo
  {author} {\bibfnamefont {T.}~\bibnamefont {Sato}}, \bibinfo {author}
  {\bibfnamefont {T.}~\bibnamefont {Katayama}}, \bibinfo {author}
  {\bibfnamefont {K.}~\bibnamefont {Ogawa}}, \bibinfo {author} {\bibfnamefont
  {H.}~\bibnamefont {Ohashi}}, \bibinfo {author} {\bibfnamefont
  {H.}~\bibnamefont {Kimura}}, \bibinfo {author} {\bibfnamefont
  {S.}~\bibnamefont {Takahashi}}, \bibinfo {author} {\bibfnamefont
  {K.}~\bibnamefont {Takeshita}}, \bibinfo {author} {\bibfnamefont
  {H.}~\bibnamefont {Tomizawa}}, \bibinfo {author} {\bibfnamefont
  {S.}~\bibnamefont {Goto}}, \bibinfo {author} {\bibfnamefont {T.}~\bibnamefont
  {Ishikawa}}, \ and\ \bibinfo {author} {\bibfnamefont {M.}~\bibnamefont
  {Yabashi}},\ }\href {\doibase 10.1088/1367-2630/15/8/083035} {\bibfield
  {journal} {\bibinfo  {journal} {New Journal of Physics}\ }\textbf {\bibinfo
  {volume} {15}},\ \bibinfo {pages} {83035} (\bibinfo {year}
  {2013})}\BibitemShut {NoStop}%
\bibitem [{\citenamefont {Hoszowska}\ \emph {et~al.}(2009)\citenamefont
  {Hoszowska}, \citenamefont {Kheifets}, \citenamefont {Dousse}, \citenamefont
  {Berset}, \citenamefont {Bray}, \citenamefont {Cao}, \citenamefont {Fennane},
  \citenamefont {Kayser}, \citenamefont {Kav{\v{c}}i{\v{c}}}, \citenamefont
  {Szlachetko},\ and\ \citenamefont {Szlachetko}}]{Hoszowska2009}%
  \BibitemOpen
  \bibfield  {author} {\bibinfo {author} {\bibfnamefont {J.}~\bibnamefont
  {Hoszowska}}, \bibinfo {author} {\bibfnamefont {A.~K.}\ \bibnamefont
  {Kheifets}}, \bibinfo {author} {\bibfnamefont {J.-C.}\ \bibnamefont
  {Dousse}}, \bibinfo {author} {\bibfnamefont {M.}~\bibnamefont {Berset}},
  \bibinfo {author} {\bibfnamefont {I.}~\bibnamefont {Bray}}, \bibinfo {author}
  {\bibfnamefont {W.}~\bibnamefont {Cao}}, \bibinfo {author} {\bibfnamefont
  {K.}~\bibnamefont {Fennane}}, \bibinfo {author} {\bibfnamefont
  {Y.}~\bibnamefont {Kayser}}, \bibinfo {author} {\bibfnamefont
  {M.}~\bibnamefont {Kav{\v{c}}i{\v{c}}}}, \bibinfo {author} {\bibfnamefont
  {J.}~\bibnamefont {Szlachetko}}, \ and\ \bibinfo {author} {\bibfnamefont
  {M.}~\bibnamefont {Szlachetko}},\ }\href {\doibase
  10.1103/PhysRevLett.102.073006} {\bibfield  {journal} {\bibinfo  {journal}
  {Physical Review Letters}\ }\textbf {\bibinfo {volume} {102}},\ \bibinfo
  {pages} {73006} (\bibinfo {year} {2009})}\BibitemShut {NoStop}%
\bibitem [{\citenamefont {Cowan}(1981)}]{cowan}%
  \BibitemOpen
  \bibfield  {author} {\bibinfo {author} {\bibfnamefont {R.}~\bibnamefont
  {Cowan}},\ }\href@noop {} {\emph {\bibinfo {title} {{The Theory of Atomic
  Structure and Spectra}}}},\ \bibinfo {edition} {1st}\ ed.,\ Los Alamos Series
  in Basic and Applied Sciences\ (\bibinfo  {publisher} {University of
  California Press},\ \bibinfo {year} {1981})\BibitemShut {NoStop}%
\bibitem [{\citenamefont {Kelly}(1976)}]{Kelly1976}%
  \BibitemOpen
  \bibfield  {author} {\bibinfo {author} {\bibfnamefont {H.~P.}\ \bibnamefont
  {Kelly}},\ }\href {\doibase 10.1103/PhysRevLett.37.386} {\bibfield  {journal}
  {\bibinfo  {journal} {Physical Review Letters}\ }\textbf {\bibinfo {volume}
  {37}},\ \bibinfo {pages} {386} (\bibinfo {year} {1976})}\BibitemShut
  {NoStop}%
\bibitem [{\citenamefont {{\AA}berg}\ \emph {et~al.}(1976)\citenamefont
  {{\AA}berg}, \citenamefont {Jamison},\ and\ \citenamefont
  {Richard}}]{Aberg1976}%
  \BibitemOpen
  \bibfield  {author} {\bibinfo {author} {\bibfnamefont {T.}~\bibnamefont
  {{\AA}berg}}, \bibinfo {author} {\bibfnamefont {K.~A.}\ \bibnamefont
  {Jamison}}, \ and\ \bibinfo {author} {\bibfnamefont {P.}~\bibnamefont
  {Richard}},\ }\href {\doibase 10.1103/PhysRevLett.37.63} {\bibfield
  {journal} {\bibinfo  {journal} {Physical Review Letters}\ }\textbf {\bibinfo
  {volume} {37}},\ \bibinfo {pages} {63} (\bibinfo {year} {1976})}\BibitemShut
  {NoStop}%
\bibitem [{\citenamefont {Hoogkamer}\ \emph {et~al.}(1976)\citenamefont
  {Hoogkamer}, \citenamefont {Woerlee}, \citenamefont {Saris},\ and\
  \citenamefont {Gavrila}}]{Hoogkamer1976}%
  \BibitemOpen
  \bibfield  {author} {\bibinfo {author} {\bibfnamefont {T.~P.}\ \bibnamefont
  {Hoogkamer}}, \bibinfo {author} {\bibfnamefont {P.}~\bibnamefont {Woerlee}},
  \bibinfo {author} {\bibfnamefont {F.~W.}\ \bibnamefont {Saris}}, \ and\
  \bibinfo {author} {\bibfnamefont {M.}~\bibnamefont {Gavrila}},\ }\href
  {\doibase 10.1088/0022-3700/9/6/004} {\bibfield  {journal} {\bibinfo
  {journal} {Journal of Physics B: Atomic and Molecular Physics}\ }\textbf
  {\bibinfo {volume} {9}},\ \bibinfo {pages} {L145} (\bibinfo {year}
  {1976})}\BibitemShut {NoStop}%
\bibitem [{\citenamefont {Hodge}(1977)}]{Hodge1977}%
  \BibitemOpen
  \bibfield  {author} {\bibinfo {author} {\bibfnamefont {B.}~\bibnamefont
  {Hodge}},\ }\href {\doibase 10.1103/PhysRevA.16.1543} {\bibfield  {journal}
  {\bibinfo  {journal} {Physical Review A}\ }\textbf {\bibinfo {volume} {16}},\
  \bibinfo {pages} {1543} (\bibinfo {year} {1977})}\BibitemShut {NoStop}%
\bibitem [{\citenamefont {W{\"{o}}lfli}\ \emph {et~al.}(1975)\citenamefont
  {W{\"{o}}lfli}, \citenamefont {Stoller}, \citenamefont {Bonani},
  \citenamefont {Suter},\ and\ \citenamefont {St{\"{o}}ckli}}]{Wolfli1975}%
  \BibitemOpen
  \bibfield  {author} {\bibinfo {author} {\bibfnamefont {W.}~\bibnamefont
  {W{\"{o}}lfli}}, \bibinfo {author} {\bibfnamefont {C.}~\bibnamefont
  {Stoller}}, \bibinfo {author} {\bibfnamefont {G.}~\bibnamefont {Bonani}},
  \bibinfo {author} {\bibfnamefont {M.}~\bibnamefont {Suter}}, \ and\ \bibinfo
  {author} {\bibfnamefont {M.}~\bibnamefont {St{\"{o}}ckli}},\ }\href {\doibase
  10.1103/PhysRevLett.35.656} {\bibfield  {journal} {\bibinfo  {journal}
  {Physical Review Letters}\ }\textbf {\bibinfo {volume} {35}},\ \bibinfo
  {pages} {656} (\bibinfo {year} {1975})}\BibitemShut {NoStop}%
\bibitem [{\citenamefont {Stoller}\ \emph {et~al.}(1977)\citenamefont
  {Stoller}, \citenamefont {W{\"{o}}lfli}, \citenamefont {Bonani},
  \citenamefont {St{\"{o}}ckli},\ and\ \citenamefont {Suter}}]{Stoller1977}%
  \BibitemOpen
  \bibfield  {author} {\bibinfo {author} {\bibfnamefont {C.}~\bibnamefont
  {Stoller}}, \bibinfo {author} {\bibfnamefont {W.}~\bibnamefont
  {W{\"{o}}lfli}}, \bibinfo {author} {\bibfnamefont {G.}~\bibnamefont
  {Bonani}}, \bibinfo {author} {\bibfnamefont {M.}~\bibnamefont
  {St{\"{o}}ckli}}, \ and\ \bibinfo {author} {\bibfnamefont {M.}~\bibnamefont
  {Suter}},\ }\href {\doibase 10.1103/PhysRevA.15.990} {\bibfield  {journal}
  {\bibinfo  {journal} {Physical Review A}\ }\textbf {\bibinfo {volume} {15}},\
  \bibinfo {pages} {990} (\bibinfo {year} {1977})}\BibitemShut {NoStop}%
\bibitem [{\citenamefont {Eisenberger}\ \emph {et~al.}(1976)\citenamefont
  {Eisenberger}, \citenamefont {Platzman},\ and\ \citenamefont
  {Winick}}]{Eisenberger1976}%
  \BibitemOpen
  \bibfield  {author} {\bibinfo {author} {\bibfnamefont {P.}~\bibnamefont
  {Eisenberger}}, \bibinfo {author} {\bibfnamefont {P.~M.}\ \bibnamefont
  {Platzman}}, \ and\ \bibinfo {author} {\bibfnamefont {H.}~\bibnamefont
  {Winick}},\ }\href {\doibase 10.1103/PhysRevB.13.2377} {\bibfield  {journal}
  {\bibinfo  {journal} {Physical Review B}\ }\textbf {\bibinfo {volume} {13}},\
  \bibinfo {pages} {2377} (\bibinfo {year} {1976})}\BibitemShut {NoStop}%
\bibitem [{\citenamefont {Goudsmit}\ and\ \citenamefont
  {Gropper}(1931)}]{Goudsmit1931}%
  \BibitemOpen
  \bibfield  {author} {\bibinfo {author} {\bibfnamefont {S.}~\bibnamefont
  {Goudsmit}}\ and\ \bibinfo {author} {\bibfnamefont {L.}~\bibnamefont
  {Gropper}},\ }\href {\doibase 10.1103/PhysRev.38.225} {\bibfield  {journal}
  {\bibinfo  {journal} {Physical Review}\ }\textbf {\bibinfo {volume} {38}},\
  \bibinfo {pages} {225} (\bibinfo {year} {1931})}\BibitemShut {NoStop}%
\bibitem [{\citenamefont {Berger}\ \emph {et~al.}()\citenamefont {Berger},
  \citenamefont {Hubbell}, \citenamefont {Seltzer}, \citenamefont {Chang},
  \citenamefont {Coursey}, \citenamefont {Sukumar}, \citenamefont {Zucker},\
  and\ \citenamefont {Olsen}}]{xcom}%
  \BibitemOpen
  \bibfield  {author} {\bibinfo {author} {\bibfnamefont {M.}~\bibnamefont
  {Berger}}, \bibinfo {author} {\bibfnamefont {J.}~\bibnamefont {Hubbell}},
  \bibinfo {author} {\bibfnamefont {S.}~\bibnamefont {Seltzer}}, \bibinfo
  {author} {\bibfnamefont {J.}~\bibnamefont {Chang}}, \bibinfo {author}
  {\bibfnamefont {J.}~\bibnamefont {Coursey}}, \bibinfo {author} {\bibfnamefont
  {R.}~\bibnamefont {Sukumar}}, \bibinfo {author} {\bibfnamefont
  {D.}~\bibnamefont {Zucker}}, \ and\ \bibinfo {author} {\bibfnamefont
  {K.}~\bibnamefont {Olsen}},\ }\href@noop {} {\bibinfo  {journal} {National
  Institute of Standards and Technology, Gaithersburg, MD}\ }\BibitemShut
  {NoStop}%
\bibitem [{\citenamefont {Deutsch}\ and\ \citenamefont
  {Kizler}(1992)}]{Deutsch1992}%
  \BibitemOpen
\bibfield  {journal} {  }\bibfield  {author} {\bibinfo {author} {\bibfnamefont
  {M.}~\bibnamefont {Deutsch}}\ and\ \bibinfo {author} {\bibfnamefont
  {P.}~\bibnamefont {Kizler}},\ }\href {\doibase 10.1103/PhysRevA.45.2112}
  {\bibfield  {journal} {\bibinfo  {journal} {Physical Review A}\ }\textbf
  {\bibinfo {volume} {45}},\ \bibinfo {pages} {2112} (\bibinfo {year}
  {1992})}\BibitemShut {NoStop}%
\bibitem [{\citenamefont {Doriese}\ \emph {et~al.}(2017)\citenamefont
  {Doriese}, \citenamefont {Abbamonte}, \citenamefont {Alpert}, \citenamefont
  {Bennett}, \citenamefont {Denison}, \citenamefont {Fang}, \citenamefont
  {Fischer}, \citenamefont {Fitzgerald}, \citenamefont {Fowler}, \citenamefont
  {Gard}, \citenamefont {Hays-Wehle}, \citenamefont {Hilton}, \citenamefont
  {Jaye}, \citenamefont {McChesney}, \citenamefont {Miaja-Avila}, \citenamefont
  {Morgan}, \citenamefont {Joe}, \citenamefont {O'Neil}, \citenamefont
  {Reintsema}, \citenamefont {Rodolakis}, \citenamefont {Schmidt},
  \citenamefont {Tatsuno}, \citenamefont {Uhlig}, \citenamefont {Vale},
  \citenamefont {Ullom},\ and\ \citenamefont {Swetz}}]{Doriese2017}%
  \BibitemOpen
  \bibfield  {author} {\bibinfo {author} {\bibfnamefont {W.~B.}\ \bibnamefont
  {Doriese}}, \bibinfo {author} {\bibfnamefont {P.}~\bibnamefont {Abbamonte}},
  \bibinfo {author} {\bibfnamefont {B.~K.}\ \bibnamefont {Alpert}}, \bibinfo
  {author} {\bibfnamefont {D.~A.}\ \bibnamefont {Bennett}}, \bibinfo {author}
  {\bibfnamefont {E.~V.}\ \bibnamefont {Denison}}, \bibinfo {author}
  {\bibfnamefont {Y.}~\bibnamefont {Fang}}, \bibinfo {author} {\bibfnamefont
  {D.~A.}\ \bibnamefont {Fischer}}, \bibinfo {author} {\bibfnamefont {C.~P.}\
  \bibnamefont {Fitzgerald}}, \bibinfo {author} {\bibfnamefont {J.~W.}\
  \bibnamefont {Fowler}}, \bibinfo {author} {\bibfnamefont {J.~D.}\
  \bibnamefont {Gard}}, \bibinfo {author} {\bibfnamefont {J.~P.}\ \bibnamefont
  {Hays-Wehle}}, \bibinfo {author} {\bibfnamefont {G.~C.}\ \bibnamefont
  {Hilton}}, \bibinfo {author} {\bibfnamefont {C.}~\bibnamefont {Jaye}},
  \bibinfo {author} {\bibfnamefont {J.~L.}\ \bibnamefont {McChesney}}, \bibinfo
  {author} {\bibfnamefont {L.}~\bibnamefont {Miaja-Avila}}, \bibinfo {author}
  {\bibfnamefont {K.~M.}\ \bibnamefont {Morgan}}, \bibinfo {author}
  {\bibfnamefont {Y.~I.}\ \bibnamefont {Joe}}, \bibinfo {author} {\bibfnamefont
  {G.~C.}\ \bibnamefont {O'Neil}}, \bibinfo {author} {\bibfnamefont {C.~D.}\
  \bibnamefont {Reintsema}}, \bibinfo {author} {\bibfnamefont {F.}~\bibnamefont
  {Rodolakis}}, \bibinfo {author} {\bibfnamefont {D.~R.}\ \bibnamefont
  {Schmidt}}, \bibinfo {author} {\bibfnamefont {H.}~\bibnamefont {Tatsuno}},
  \bibinfo {author} {\bibfnamefont {J.}~\bibnamefont {Uhlig}}, \bibinfo
  {author} {\bibfnamefont {L.~R.}\ \bibnamefont {Vale}}, \bibinfo {author}
  {\bibfnamefont {J.~N.}\ \bibnamefont {Ullom}}, \ and\ \bibinfo {author}
  {\bibfnamefont {D.~S.}\ \bibnamefont {Swetz}},\ }\href {\doibase
  10.1063/1.4983316} {\bibfield  {journal} {\bibinfo  {journal} {Review of
  Scientific Instruments}\ }\textbf {\bibinfo {volume} {88}},\ \bibinfo {pages}
  {53108} (\bibinfo {year} {2017})}\BibitemShut {NoStop}%
\bibitem [{\citenamefont {Schoenlein}\ \emph {et~al.}(2007)\citenamefont
  {Schoenlein}, \citenamefont {Boutet}, \citenamefont {Minitti},\ and\
  \citenamefont {Dunne}}]{LCLS2}%
  \BibitemOpen
  \bibfield  {author} {\bibinfo {author} {\bibfnamefont {R.}~\bibnamefont
  {Schoenlein}}, \bibinfo {author} {\bibfnamefont {S.}~\bibnamefont {Boutet}},
  \bibinfo {author} {\bibfnamefont {M.}~\bibnamefont {Minitti}}, \ and\
  \bibinfo {author} {\bibfnamefont {M.}~\bibnamefont {Dunne}},\ }\href@noop {}
  {\bibfield  {journal} {\bibinfo  {journal} {Appl. Sci.}\ }\textbf {\bibinfo
  {volume} {7}},\ \bibinfo {pages} {850} (\bibinfo {year} {2007})}\BibitemShut
  {NoStop}%
\bibitem [{\citenamefont {Ament}\ \emph {et~al.}(2011)\citenamefont {Ament},
  \citenamefont {van Veenendaal}, \citenamefont {Devereaux}, \citenamefont
  {Hill},\ and\ \citenamefont {van~den Brink}}]{RIXSRMP}%
  \BibitemOpen
  \bibfield  {author} {\bibinfo {author} {\bibfnamefont {L.~J.~P.}\
  \bibnamefont {Ament}}, \bibinfo {author} {\bibfnamefont {M.}~\bibnamefont
  {van Veenendaal}}, \bibinfo {author} {\bibfnamefont {T.~P.}\ \bibnamefont
  {Devereaux}}, \bibinfo {author} {\bibfnamefont {J.~P.}\ \bibnamefont {Hill}},
  \ and\ \bibinfo {author} {\bibfnamefont {J.}~\bibnamefont {van~den Brink}},\
  }\href {\doibase 10.1103/RevModPhys.83.705} {\bibfield  {journal} {\bibinfo
  {journal} {Rev. Mod. Phys.}\ }\textbf {\bibinfo {volume} {83}},\ \bibinfo
  {pages} {705} (\bibinfo {year} {2011})}\BibitemShut {NoStop}%
\end{thebibliography}%


%
\end{document}